\documentclass[12pt,letterpaper]{article}

\usepackage{amsmath}
\usepackage[psamsfonts]{amssymb}
\usepackage{amsthm}
\usepackage{indentfirst}
\usepackage{xspace}
\usepackage{enumerate}
\usepackage{color}
\usepackage{graphicx}
\usepackage{pstricks}
\usepackage{fullpage}

\numberwithin{equation}{section}

\newcommand{\for}{\quad \text{for }}

\newcommand{\HH}{{\mathcal H}}
\newcommand{\LL}{{\mathcal L}}
\newcommand{\OO}{{\mathcal O}}
\newcommand{\cR}{{\mathcal R}}
\newcommand{\cP}{{\mathcal P}}
\newcommand{\cS}{{\mathcal S}}
\newcommand{\cQ}{{\mathcal Q}}
\newcommand{\cA}{{\mathcal A}}
\newcommand{\cB}{{\mathcal B}}
\newcommand{\cD}{{\mathcal D}}
\newcommand{\R}{\ensuremath{\mathbb R}}
\newcommand{\C}{\ensuremath{\mathbb C}}
\newcommand{\PP}{\ensuremath{\mathbb P}}
\newcommand{\Z}{\ensuremath{\mathbb Z}}

\newcommand{\half}{\ensuremath{\frac{1}{2}}}

\newcommand{\N}{{\mathcal N}}

\newcommand{\Tr}{\mathrm{Tr}}
\newcommand{\Top}{{\mathrm{top}}}
\newcommand{\YM}{{\mathrm{YM}}}
\newcommand{\BPS}{{\mathrm{BPS}}}
\newcommand{\BH}{{\mathrm{BH}}}
\newcommand{\open}{{\mathrm{open}}}
\newcommand{\im}{{\mathrm{Im}}\ }
\newcommand{\re}{{\mathrm{Re}}\ }
\newcommand{\CS}{{\mathrm{CS}}}
\newcommand{\inst}{{\mathrm{inst}}}

\newcommand{\abs}[1]{\lvert#1\rvert}
\newcommand{\norm}[1]{\lVert#1\rVert}
\newcommand{\IP}[1]{\langle#1\rangle}
\newcommand{\ket}[1]{\lvert#1\rangle}

\newcommand{\schur}{s}
\newcommand{\ghost}{\mathrm g}
\newcommand{\antighost}{\mathrm a}

\newcommand{\kahler}{K\"ahler\xspace}
\newcommand{\bo}{{\mathbf 1}}

\newcommand{\ti}[1]{\textit{#1}}

\newcommand{\fig}[3]{
\begin{figure}[t]
\begin{center}
\includegraphics[#3]{#1}
\end{center}
\caption{#2}
\label{#1}
\end{figure}
}

\begin{document}

\bibliographystyle{utphys}

\renewcommand{\baselinestretch}{1.25}
\small\normalsize

\begin{titlepage}

\begin{center}
\hfill HUTP-05/A0016\\
\hfill UCB-PTH-05/09\\
\hfill hep-th/0504057

\vskip 1.5 cm
{\huge BPS Microstates and the Open}
\vskip 0.4 cm
{\huge Topological String Wave Function}
\vskip 1.3 cm
{\large Mina Aganagic$^1$, Andrew Neitzke$^2$ and Cumrun Vafa$^2$}\\
\vskip 0.4 cm
{$^1$ University of California, Berkeley, CA 94720, USA}
\vskip 0.4 cm
{$^2$ Jefferson Physical Laboratory, Harvard University, Cambridge, MA 02138, USA}
\end{center}

\vskip 0.5 cm
\begin{abstract}
It has recently been conjectured that the closed topological string wave function
computes a grand canonical partition function 
of BPS black hole states in 4 dimensions:  $Z_{\BH}=\abs{\psi_{\Top}}^2$.  We conjecture
that the open topological string wave function also computes a
grand canonical partition function, which sums over black holes bound to BPS 
excitations on D-branes
wrapping cycles of the internal Calabi-Yau:  $Z^\open_{\BPS}=\abs{\psi^{\open}_{\Top}}^2$. 
This conjecture is verified in the case of
Type IIA on a local Calabi-Yau threefold involving a Riemann surface, where
the degeneracies of BPS states can be computed in $q$-deformed 2-dimensional Yang-Mills theory.
\end{abstract}

\end{titlepage}

\setcounter{page}{1}
\pagestyle{plain}


\tableofcontents

\section{Introduction}

The connection between topological strings and 4-dimensional BPS black holes
has been studied in recent years \cite{LopesCardoso:1998wt,LopesCardoso:1999cv,LopesCardoso:1999ur}, 
leading to a conjecture
\cite{Ooguri:2004zv} that identifies the mixed grand canonical partition function
of BPS black hole states with the squared norm of the topological string wave function: 
$Z_\BH=\abs{\psi_{\Top}}^2$.  This conjecture has been checked for certain
Calabi-Yau threefolds \cite{Vafa:2004qa,Aganagic:2004js,bh-atish}; see also the
recent related work \cite{Sen:2005pu,Dabholkar:2005by,Sen:2005ch}.
It is natural to ask how the conjecture generalizes to the case of
open topological strings.  Our primary aim in this paper is to advance
a conjecture about what the open topological string counts, and
to check it in the case of certain non-compact Calabi-Yau spaces.

We will mainly concentrate on the Type IIA superstring (and correspondingly the 
topological A model) on a non-compact Calabi-Yau threefold.  
In the closed string context, one defines the mixed black hole ensemble 
by fixing the number of D4 and D6-branes (magnetic charges) while summing over
all possible numbers of D2 and D0-branes bound to them (electric charges), weighed by chemical potentials;
this was the setup investigated in \cite{Vafa:2004qa,Aganagic:2004js}.  In our case
the Type IIA background will additionally include a finite number of
``background'' D4-branes, which wrap Lagrangian 3-cycles
of the Calabi-Yau and fill a 1+1 dimensional
subspace of Minkowski spacetime.  
In the presence of these background D4-branes one gets a gauge theory in 1+1 dimensions, 
containing new BPS states.  The role of the electric charges is played by
open D2-branes, wrapped on holomorphic discs ending on the Lagrangian 3-cycles,
while the magnetic charges are domain walls in the 1+1 dimensional theory.
We conjecture that the full topological string amplitude, including
contributions from open strings, is counting degeneracies of these BPS
states, bound to D6, D4, D2 and D0-branes:
\begin{equation}
Z_{\BPS}^\open=\abs{\psi_{\Top}^{\open}}^2.
\end{equation} 
Here $Z_\BPS^\open$ is the partition function of a mixed grand 
canonical ensemble; in this ensemble the D6 and D4-brane charges, as
well as the domain wall charge, are fixed
(and related to the real part of the topological string moduli), while
chemical potentials are turned on for the D2 and D0-branes 
(giving the imaginary parts of the moduli), including the open
D2-branes.

Our proposal is necessarily more tentative than the one given in \cite{Ooguri:2004zv}, because
one of the major planks supporting the conjecture there is missing here:  
the large-charge macroscopic/gravitational description of the BPS states we are counting has
not been studied, nor has the analogue of the attractor mechanism for these states, so we 
do not even have a classical derivation of the entropy.  Further investigations in this
direction would be extremely useful to check our conjecture.

Although we do not understand the macroscopic description of these BPS states, we can still compare
$\abs{\psi_\Top^\open}^2$ to a partition function computed from their microscopic description, 
in cases where such a description is available.
In this paper we use such a description
to check our proposal on a particular non-compact
Calabi-Yau space supporting a compact Riemann surface.  This
case was previously discussed in \cite{Vafa:2004qa,Aganagic:2004js} where the closed string conjecture was
verified.  We find that our conjecture also holds in this case.

The organization of this paper is as follows.  In Section \ref{sec-closed} we review the conjecture
in the closed string case and review its confirmation in the context of local
Riemann surfaces inside a Calabi-Yau.  In Section \ref{sec-closed-revisited} 
we explain the unexpected appearance of open topological
string amplitudes in \cite{Aganagic:2004js}, reinterpreting them in terms of purely
closed topological strings along the lines of the original conjecture
\cite{Ooguri:2004zv}.  
In Section \ref{sec-qm-open} we discuss the wave function nature of the open topological string.
In Section \ref{sec-open} 
we introduce additional branes in our physical string background and state 
our main conjecture.  In Section \ref{sec-2dym-open} we check the conjecture 
in the context of a local Calabi-Yau geometry near a Riemann surface with Lagrangian D-branes included.
Most of the computations are relegated to the appendices:  In Appendix \ref{app-group-theory} we fix some
group theory conventions and review some basic group theory facts.  In Appendix
\ref{app-qym} we review the $q$-deformed Yang-Mills theory in 2 dimensions and the computation
of its amplitudes by gluing, including insertion of some eigenvalue freezing
operators important for this paper.  In Appendix \ref{app-disc-wave-function} we express the
wave function of $q$-deformed 2d Yang-Mills on the disc in terms of theta functions.  
Finally, Appendix \ref{app-factorization} discusses
many issues related to the large $N$ limit of our computations, and 
the factorization of the BPS partition function at large $N$ in terms of
topological and anti-topological contributions.  In particular, we give a physical
explanation of the factorization of the $q$-deformed Yang-Mills amplitudes
in the large $N$ limit.

\section{The closed string case} \label{sec-closed}

In \cite{Ooguri:2004zv} a duality was conjectured which relates counting of
microstates of supersymmetric black holes which arise in compactification of type
II string theory on a Calabi-Yau threefold $X$ and closed topological string
theory on $X$.  In this section we review this conjecture and one case in which
it has been explicitly checked.

Consider Type IIA on $X \times \R^{3,1}$.  
One can obtain charged BPS black holes 
in $\R^{3,1}$ by wrapping D6, D4, D2 and D0-branes over holomorphic cycles in $X$.
The charges of the black hole are determined by the choice of holomorphic cycles; the
intersection pairing in $X$ gives rise to the electric-magnetic pairing in $\R^{3,1}$,
and we refer to D6 and D4-brane charges as ``magnetic'' while D2 and D0-brane
charges are ``electric.''  Then one can define a mixed ensemble of BPS black hole states by
fixing the D6 and D4-brane charges $Q_6$, $Q_4$, and summing over D2 and D0-brane
charges with fixed chemical potentials $\varphi_2$, $\varphi_0$.  One can write a partition function 
for this ensemble,
\begin{equation} \label{bh-partition-function}
Z_\BH(Q_6, Q_4, \varphi_2, \varphi_0) = \sum_{Q_2, Q_0} \Omega_{Q_6, Q_4, Q_2, Q_0} e^{-Q_2 \varphi_2 - Q_0 \varphi_0}.
\end{equation} 
Here $\Omega_{Q_6, Q_4, Q_2, Q_0}$ is the contribution from BPS bound states
with fixed D-brane charge.  

The conjecture of \cite{Ooguri:2004zv} is that
\begin{equation} \label{osv-conjecture}
Z_\BH(Q_6, Q_4, \varphi_2, \varphi_0) = \abs{\psi_\Top(g_\Top, t)}^2,
\end{equation}
where $\psi_\Top(g_\Top, t)$ denotes the A model topological string partition function, evaluated
at the topological string coupling\footnote{$Q_4$ is naturally a class in $H_4(X,\Z)$, which we are relating
to $t \in H^2(X,\C)$, and $Q_6$ is naturally a class in $H_6(X,\Z)$, which we
are relating to $H^0(X,\C) = \C$.}
\begin{equation} \label{osv-coupling}
g_\Top = \frac{4 \pi i}{i \frac{\varphi_0}{\pi} + Q_6},
\end{equation}
and \kahler parameter
\begin{equation} \label{osv-kahler}
t = \half g_\Top \left(i \frac{\varphi_2}{\pi} + Q_4 \right).
\end{equation}
The real parts of the parameters \eqref{osv-coupling} and \eqref{osv-kahler} are dictated by
the ``attractor mechanism'' of $\N=2, d=4$ supergravity \cite{Ferrara:1995ih,Strominger:1996kf},
which relates the moduli of $X$ near a black hole horizon to the black hole charges.

One can (at least formally) invert the relation \eqref{osv-conjecture} to recover the microcanonical
degeneracies $\Omega$ from $\abs{\psi_\Top}^2$, via the integral formula
\begin{equation} \label{wigner-transform}
\Omega_{Q_6, Q_4, Q_2, Q_0} = \int d\varphi_2\, d\varphi_0\, e^{Q_0 \varphi_0 + Q_2 \varphi_2} \abs{\psi_\Top}^2.
\end{equation} 
This formula has a natural interpretation from the point of view of the wave function interpretation
of $\psi_\Top$ developed in \cite{Witten:1993ed,Dijkgraaf:2002ac}
as an interpretation of the holomorphic anomaly
\cite{Bershadsky:1993ta,Bershadsky:1994cx}. 
Namely, \eqref{wigner-transform}
expresses $\Omega$ as the ``Wigner function'' (phase-space density) associated to 
$\psi_\Top$.  The background-independent generalization of this transform and its relation
to the counting of black hole states has been further elucidated in \cite{Verlinde:2004ck}.

The formula \eqref{wigner-transform} also illustrates a crucial point about the conjecture:  in order to
use it to compute $\Omega$, one would need to know the full $\abs{\psi_\Top}^2$, not only 
its asymptotic expansion for $g_\Top \ll 1$.  Put another way, knowing the BPS degeneracies $\Omega$ is in
some sense equivalent to having a nonperturbative completion of $\abs{\psi_\Top}^2$.

\subsection*{A solvable example}

In this section we review the work of \cite{Vafa:2004qa,Aganagic:2004js}
which argued that the conjecture \eqref{osv-conjecture} holds
in the case where $X$ is a particular non-compact Calabi-Yau threefold, namely the total space of a
holomorphic vector bundle over a compact Riemann surface $\Sigma$
of genus $g$,
\begin{equation}
X = \LL_{-p} \oplus \LL_{p+2g-2} \to \Sigma,
\end{equation}
for some $p>0$.\footnote{By $\LL_k$ we mean a holomorphic line bundle of degree
$k$ over $\Sigma$.}

The idea is that for this $X$ one can use 2-dimensional Yang-Mills theory to compute
$Z_\BH$, as follows.  Suppose we wrap $N$ D4-branes on the holomorphic 4-cycle
\begin{equation}
\cD = \LL_{-p} \to \Sigma.
\end{equation}
Then the theory on the D4-branes (in the Calabi-Yau directions) is the
$\N=4$ supersymmetric Yang-Mills theory, or more precisely a
topologically twisted version of that theory, as explained in \cite{Bershadsky:1996qy}.  
The path integral in this theory includes configurations in
which D0-branes, and D2-branes wrapping $\Sigma$, 
are bound to the D4-branes.  
Hence the partition function of the 4-dimensional twisted supersymmetric 
gauge theory computes
a sum over the mixed ensemble of BPS states which we considered above.  The 
D4 and D6-brane charges are 
\begin{align}
Q_4 &= N [\cD], \\
Q_6 &= 0.
\end{align}
The chemical potentials for the brane charges are roughly given by the masses of the
branes (for the D2-branes we turn on a Ramond-Ramond field $\theta$):
\begin{align}
\varphi_0 &= 4 \pi^2 / g_s, \\
\varphi_2 &= 2 \pi p \theta / g_s.
\end{align}
Since the gauge theory sums over all brane charges we can now 
write\footnote{There are some subtleties because of the non-compactness of $X$, 
as noted in \cite{Aganagic:2004js}:  $Z_\YM$ turns out to give
a sum over finitely many sectors, each with a $g_s$-dependent prefactor.}
\begin{equation}
Z_\YM = Z_\BH.
\end{equation}
It was argued in \cite{Vafa:2004qa} that, for the purpose of computing $Z_\YM$,
we can restrict to field configurations in the $\N=4$ theory which are invariant 
under the $U(1)$ action on the fibers of $\LL_{-p}$.  One then
obtains $Z_\YM$ as the partition function of a $q$-deformed Yang-Mills theory on $\Sigma$ (see Appendix
\ref{app-qym}), where $\Sigma$ has area $p$ and the parameters are fixed by
\begin{equation} \label{2dym-params}
\theta_{\YM} = \theta, \quad g^2_{\YM} = g_s, \quad q = e^{-g_s}.
\end{equation}
The $q$-deformed Yang-Mills theory is a relative of the ordinary Yang-Mills theory
in two dimensions, and shares with that theory the property of being exactly solvable; the
topological string on $X$ is also exactly solvable to all orders in perturbation theory (using
recent results of \cite{Bryan:2004iq} in the case $g > 1$). 
Hence we can use $X$ as a testing ground for \eqref{osv-conjecture}.
More precisely, since we do not have a good
understanding of the nonperturbative topological string, what we can do is
look at the asymptotic expansion of $\abs{\psi_\Top}^2$ in the limit $g_s \ll 1$, with $t$ fixed.
On the physical side this corresponds to taking $\varphi_0$, $\varphi_2$, and $N$ to infinity with
fixed ratios (this is a 't Hooft limit in the Yang-Mills theory.)

In this limit one finds that $Z_\YM$ factorizes into a sum of ``conformal blocks,''
each given by the topological string on $X$, with some D-branes inserted as we will
explain below:
\begin{multline} \label{2dym-factorization}
Z_\YM(\varphi_0, \varphi_2, N) = \\
\sum_{R'_1, \dots, R'_{\abs{2g-2}}}  \sum_{l \in \Z} 
\psi_\Top^{R'_1, \dots, R'_{\abs{2g-2}}} (g_\Top, t + l p g_\Top) \overline{\psi_\Top^{R'_1, \dots, R'_{\abs{2g-2}}}(g_\Top, t - l p g_\Top)} + \OO(e^{-N}).
\end{multline}
Here $t$ and $g_\Top$ are as dictated by \eqref{osv-coupling} and \eqref{osv-kahler},
namely,
\begin{align}
g_\Top &= 4 \pi^2 / \varphi_0 = g_s, \label{osv-coupling-solvable} \\
t & = \half g_\Top \left(\#(\Sigma \cap \cD) N + i \varphi_2 / \pi \right) = \half N (p+2g-2) g_s + i p \theta. \label{osv-kahler-solvable}
\end{align}
The index $l$ was interpreted in \cite{Vafa:2004qa} as measuring the Ramond-Ramond
flux through $\Sigma$.
The labels $R'_i$ are subtler; they appear only when $g \neq 1$, in which case they were
interpreted in \cite{Aganagic:2004js} as running over boundary conditions on $\abs{2g-2}$ infinite
stacks of D-branes (which we christen ``ghost'' D-branes) 
in the topological string.  Each stack lies on a Lagrangian submanifold of $X$, intersecting $\cD$
in an $S^1$ in the fiber of $\LL_{p+2g-2}$ over a point.
The boundary conditions on each stack are specified by a choice
of a Young diagram $R'$.\footnote{All primed quantities which appear in this paper
are associated to these ghost D-branes.}

The form of \eqref{2dym-factorization} looks different from that of \eqref{osv-conjecture}.  Nevertheless, as we will explain in the next section, the sum over Young diagrams $R'_i$
is indeed consistent with \eqref{osv-conjecture}, when we take 
into account extra closed string moduli at infinity.

\section{Revisiting the closed string theory} \label{sec-closed-revisited}

In this section we revisit the relation between 2-d Yang-Mills theory and 
the closed topological string, with the aim of giving a better 
interpretation to the sum over chiral blocks and the appearance of
``ghost'' D-branes.

As we reviewed in Section \ref{sec-closed}, the partition function of the twisted $U(N)$ Yang-Mills theory on
$\cD = {\cal L}_{-p} \rightarrow \Sigma$ factorizes at large $N$ as a sum of blocks, each of which can be 
interpreted as the square of a topological string amplitude involving $2g-2$ infinite stacks of ghost branes.  Introducing 
a $U(\infty)$-valued holonomy $U'_i = e^{u'_i}$ on each stack of ghost branes, we can rewrite \eqref{2dym-factorization}
as
\begin{equation}\label{2dym-factorization-integralform}
Z_\YM = \sum_{l \in \Z} \int d_H u'_1 \cdots d_H u'_{2g-2} \; \psi^\ghost_\Top(g_\Top, u', t + l p g_\Top) \overline{\psi^\ghost_\Top(g_\Top, u', t - l p g_\Top)},
\end{equation}
where
\begin{equation}\label{ztop-ghost-holonomies}
\psi^\ghost_\Top(g_\Top, u', t) = 
\sum_{R'_1,\ldots R'_{2g-2}} \psi_\Top^{R'_1,\ldots, R'_{2g-2}}(g_\Top, t)
e^{-\half N g_s \sum_{i=1}^{2g-2} \abs{R'_i}} \prod_{i=1}^{2g-2} \schur_{R'_i}(e^{u'_i}).
\end{equation}
For $g = 0$ the formula is similar, except that the role of ghost branes and ghost antibranes
are reversed in the antitopological amplitude:
\begin{equation}\label{2dym-factorization-integralform-g0}
Z_\YM = \sum_{l \in \Z} \int d_H u'_1 d_H u'_2 \; \psi^\ghost_\Top(g_\Top, u', t + l p g_\Top) \overline{\psi^\antighost_\Top(g_\Top, u', t - l p g_\Top)},
\end{equation}
where
\begin{align}\label{ztop-ghost-holonomies-g0}
\psi^\ghost_\Top(g_\Top, u', t) &= \sum_{R'_1, R'_2} \psi_\Top^{R'_1, R'_2}(g_\Top, t) e^{-\half N g_s (\abs{R'_1} + \abs{R'_2})} \schur_{R'_1} (e^{u'_1}) \schur_{R'_2}(e^{u'_2}), \\
\psi^\antighost_\Top(g_\Top, u', t) &= \sum_{R'_1, R'_2} (-)^{\abs{R'_1}+\abs{R'_2}} \psi_\Top^{R'_1, R'_2}(g_\Top, t) e^{-\half N g_s (\abs{R'_1} + \abs{R'_2})} \schur_{R'^t_1} (e^{u'_1}) \schur_{R'^t_2}(e^{u'_2}).
\end{align}
The change from branes to antibranes is reflected in the signs $(-)^{\abs{R'}}$ and the switch $R' \to R'^t$ between
$\psi^\ghost$ and $\psi^\antighost$, as in \cite{Aganagic:2003db}.

Now note that \eqref{2dym-factorization-integralform} and \eqref{2dym-factorization-integralform-g0} look like the integral 
\eqref{wigner-transform}, that computes the microcanonical degeneracies by integrating over
the imaginary part of each \kahler modulus while the real part is fixed by the corresponding magnetic
charge.  Indeed, the factor $e^{-\half N g_s \sum_{i=1}^{2g-2} \abs{R'_i}}$ could be 
absorbed in $U'$, at the expense of making it non-unitary:
this just amounts to giving $u'$ a real part.
This is reminiscent of the ``attractor'' formula \eqref{osv-kahler}, which says the real part of the \kahler modulus
is related to the charge.  So indeed, \eqref{2dym-factorization-integralform} could be consistent with 
the conjecture \eqref{osv-conjecture}, if we somehow regard $u'$ as an extra closed string modulus;
then there would be electric and magnetic charges corresponding to it, and \eqref{2dym-factorization-integralform}
says that $Z_\YM$ is the partition function of an ensemble in which we have fixed these charges.  
As we will now explain, this interpretation of $u'$ is indeed plausible.

\subsection*{Open vs. closed}

We explained above that the nonperturbative completion of the closed
topological string appears to involve Lagrangian D-branes on the
Calabi-Yau manifold. The appearance of open string amplitudes in
this context is surprising, since in the physical string this 
would have half as much supersymmetry as we have available.  
As we will now argue, the correct interpretation involves not open but closed strings.

Namely, as was shown in \cite{top-integ}, in the
topological string, inserting non-compact D-branes is equivalent
to turning on certain non-normalizable deformations of the
Calabi-Yau.  This is an open-closed duality of the
topological string, generalizing the well-known duality for
D-branes on compact cycles.  This means that, at the level of the
topological string, we can interpret the modulus $U'$ in
\eqref{2dym-factorization-integralform} as either corresponding to an open string
configuration or to a boundary condition at infinity
of the closed topological string.  In the physical string theory,
however, we do not have this freedom; since there are no
Ramond-Ramond fluxes turned on, the only interpretation available
is the closed string one. 

The torus symmetries of the Calabi-Yau manifold can be used to constrain the
types of deformation that we consider.  Namely, the Lagrangian D-branes to which
\eqref{ztop-ghost-holonomies} corresponds respect the torus symmetries,
and the gravitational backreaction they create does so as well.
Such torus invariant deformations, normalizable and not, were studied in \cite{top-integ},
so we can borrow the results of that paper.
The topological string theory in \cite{top-integ} was described as
the theory of a chiral boson on a Riemann surface,
and the Lagrangian D-branes were coherent states of this chiral boson. 
(Note here that we are using the mirror B-model language.
The global action of mirror symmetry on $X$ is not relevant for us;
this is merely a convenient language in which to describe the behavior near an asymptotic
infinity.)  The non-normalizable deformations of the
Calabi-Yau near an asymptotic infinity\footnote{In the cases studied in \cite{top-integ} there is a clear notion
of what ``an asymptotic infinity'' means:  it means a toric 2-cycle which extends to infinity.  In the cases we are considering
here the situation is not as rigorously understood, but we will make some comments below.} 
can be parameterized by the coherent states of the chiral boson:
\begin{equation}
\ket{\tau} = \exp\left(\sum_{n>0} \tau_n {\alpha_{-n}}\right) \ket{0},
\end{equation} 
where $\alpha_n$ are the chiral boson creation and
annihilation operators. 

The parameters $\tau$ are related to the D-brane holonomies by
\begin{equation}\label{iden}
\tau_n = g_s \Tr U'^n,
\end{equation}
where $\Tr$ denotes the trace in the fundamental representation, The
factor of $g_s$ is needed to convert an open string amplitude in terms
of $U$ to a closed string amplitude in terms of $t$; it appears
because a trace of $U$ in the fundamental representation couples to a
hole in the string worldsheet, and the hole is in turn weighted by
$g_s$ in the string perturbation expansion.
In this sense the open string modulus $U$ can be traded for the infinite collection of closed string moduli $\tau_n$.  

Actually, it is more convenient to reparameterize slightly by taking a logarithm, writing
$\tau_n = e^{-t_n}$.  The point is that the A model partition function turns out to be an expansion in
$e^{-t_n}$, so the moduli $t_n$ appear on the same footing as the \kahler volumes $t$ of compact cycles.
Indeed, we can think of them as representing \kahler volumes of classes in $H_2(X, \Z)$ (with
some appropriate notion of what $H_2(X,\Z)$ means for this non-compact $X$.)
What can we say about these classes?
In the cases considered in \cite{top-integ}, for each asymptotic infinity there is a holomorphic disc $C$ which ``ends'' on it,
and $t_n$ represents a class which contains $n[C]$ as well as some extra contributions at infinity.
In the open string language, the disc $C$ can be thought of as ending on the Lagrangian
branes which represent the deformations at this asymptotic infinity.

\subsection*{The attractor mechanism and ghost D-branes}

Now we come to the interpretation of the shift $U' \to U' e^{-\half N g_s}$, or equivalently
\begin{equation} \label{t-shift}
\re t_n = \half n N g_s.
\end{equation}
Such shifts have frequently appeared in the topological string in the presence of D-branes.
Here we can understand the shift as a reflection of the attractor mechanism on the
closed string moduli.  
Namely, in the case we are considering here, $C$ is a disc in the fiber of $\LL_{p+2g-2}$, which 
intersects $\cD$ at one point, as shown in Figure \ref{fig-fiber-disc-outline-fonts-small}.
\fig{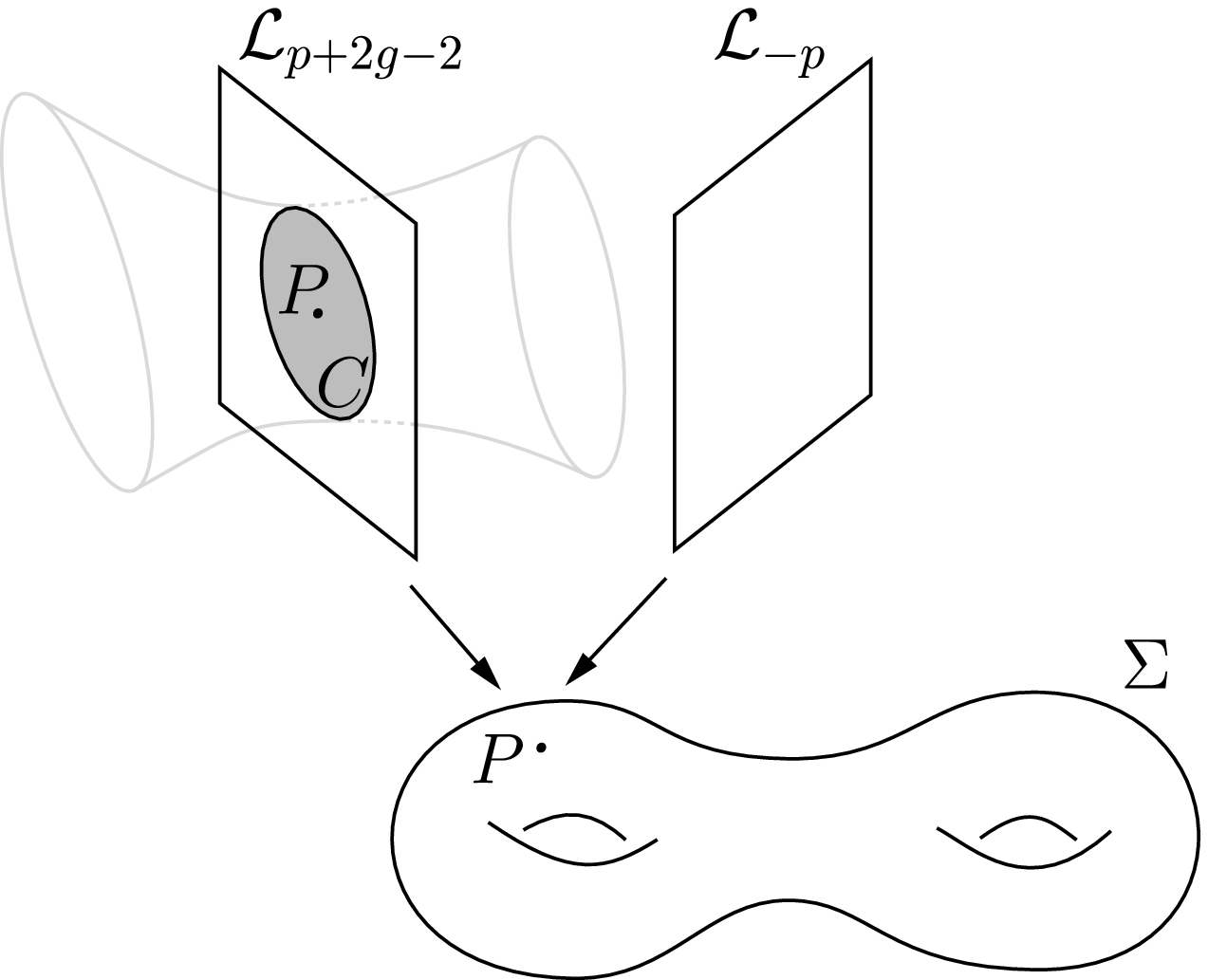}{The disc $C$ in the fiber of $\LL_{p+2g-2}$ over a point $P$ on the Riemann surface $\Sigma$;
$C$ meets $\cD$ only at $P$, and the boundary of $C$ lies on the Lagrangian submanifold representing
this asymptotic infinity.}{height=2.4in}
Then $\half n N g_s$ is exactly the expected attractor value for the \kahler 
modulus $t_n$, as follows from \eqref{osv-kahler}, the fact that the D4-brane 
charge is $Q_4 = N[\cD]$, and $\#(C \cap \cD) = 1$.  (Whatever the extra contributions
at infinity to the class represented by $t_n$ are, they have zero intersection number
with $\cD$, so they do not affect the attractor modulus.)

\subsection*{Why $2g-2$ asymptotic infinities}

The discussion of the last few sections raises
a natural question:  why are there precisely $\abs{2g-2}$
asymptotic infinities on $X$ where we can have deformations?

In general we should have expected that in a non-compact Calabi-Yau
we should include some closed string moduli coming from infinity.
However, in problems with symmetries, it is natural to conjecture that
the only relevant extra moduli from infinity are invariant under
the corresponding symmetries.  We will assume this here, and look
for symmetries in our problem which simplify the task of specifying
the closed string moduli coming from infinity.

A priori, one might have expected boundary moduli 
associated to the $\C^2$ fiber over each point of the Riemann surface.  Here
we have in addition D4-branes wrapping a line bundle
over the Riemann surface.  We claim that this implies that 
effectively we should view that direction as ``compact,'' 
or more precisely, we should view it as a degenerate limit of
a compact 4-cycle.  
After this reduction, we would expect to find boundary
moduli corresponding to a $\C$ fiber over each
point on the Riemann surface.

However, there are symmetries of the problem coming from meromorphic
vector fields on the Riemann surface.  Hence the variation of the data
at infinity can be localized at poles or zeroes of such a vector field
(deleting these points would give a well defined free action).
A generic holomorphic vector field on a Riemann surface of genus $g>1$ is nonvanishing
and well defined away from $2g-2$ poles,
which we identify with places where the asymptotic boundary condition
at infinity can be localized.  The local picture is as shown in Figure \ref{fig-asymptotic-infinity-outline-fonts-small}.
\fig{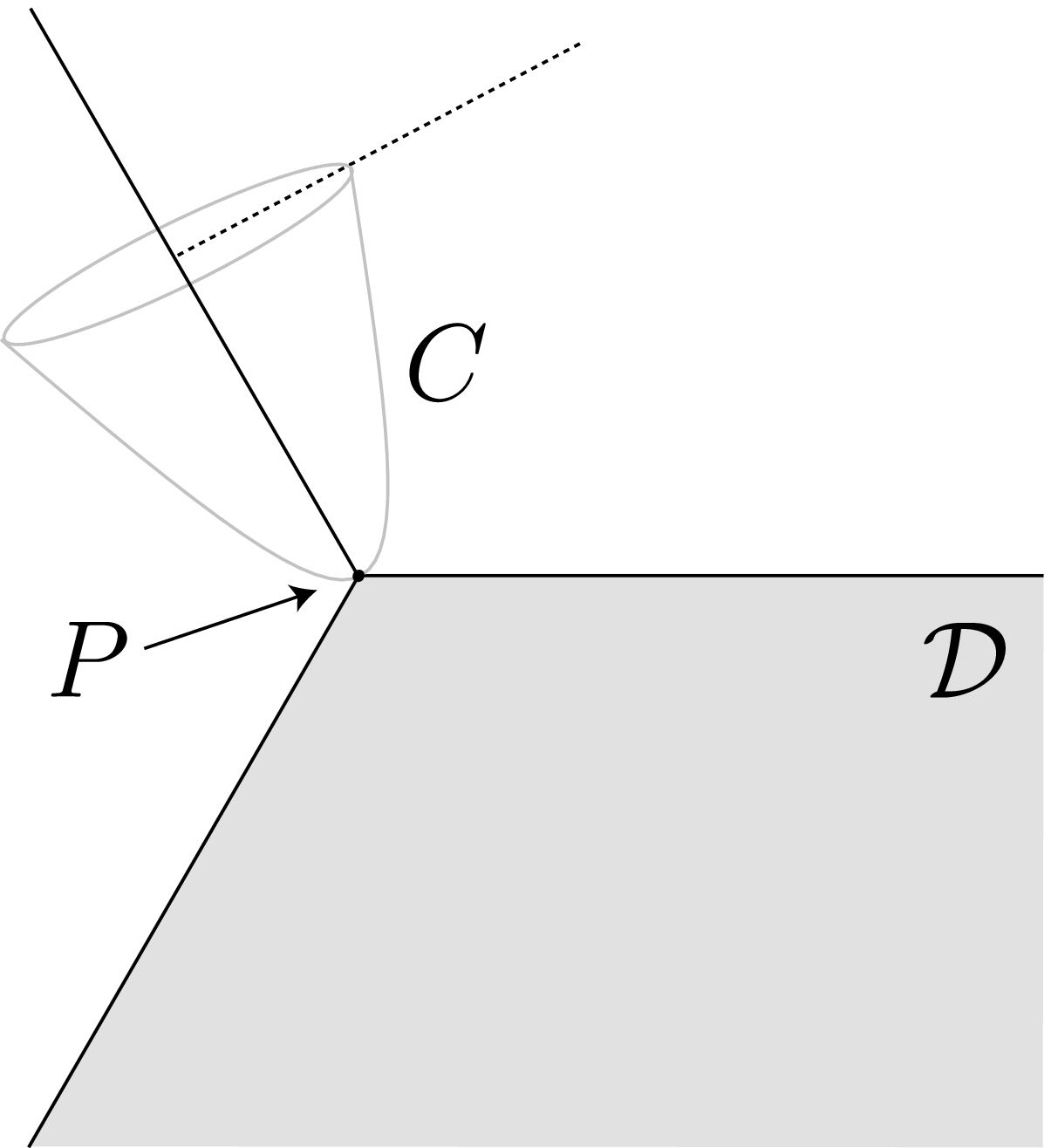}{A rough toric representation of the behavior of $X$ in a neighborhood of a singularity of the vector field $v$ described in the text.  Two of the three
$U(1)$ actions making up the toric fiber are the rotations of the line bundles ${\mathcal L}_{-p} \oplus {\mathcal L}_{p+2g-2}$ and the third is the action of $v$.  The toric base of the divisor $\cD$ on which the D4-branes are wrapped is indicated, as is the base of the Lagrangian submanifold representing the asymptotic infinity.  The disc $C$ ends on this Lagrangian submanifold, meeting $\cD$ at the single
point $P$.}{height=2.5in}

So the closed string moduli at these $2g-2$ asymptotic infinities may be identified with
the ``ghost D-brane'' contributions, as discussed above.
In the case of genus 1 there are no fixed points, which is consistent
with the fact that no ghost D-branes were needed in this case.  For genus 0 we have
a holomorphic vector field with 2 zeroes, which again suggests that we can localize
the contribution from infinity at 2 points.

This is a heuristic argument, but we feel that it captures the correct physics.

\section{The quantum mechanics of open strings} \label{sec-qm-open}

In Section \ref{sec-closed} we reviewed the conjecture of \cite{Ooguri:2004zv}
and its relation to the wave function nature of the closed topological string.
In this section we recall the parallel statement for the open topological string.  
The fact that the open topological string 
partition function including non-compact branes is a wave function 
was first noticed in \cite{top-integ}, and was crucial in that paper for the 
solution of the B model.  In this section we will give two ways of understanding this
wave function property:  a direct route via canonical quantization of Chern-Simons
theory, and a more indirect one via the holomorphic anomaly (background dependence)
for open strings.

\subsection*{Canonical quantization in Chern-Simons}

Recall that the topological A model string theory on 
$M$ D-branes wrapped on a Lagrangian cycle $L$ 
is the $U(M)$ Chern-Simons theory deformed
by worldsheet instanton corrections:
\begin{equation}
S = S_\CS +S_\inst,
\end{equation} 
where 
\begin{equation}
S_\CS = \frac{4\pi i}{k} \int_L \Tr \left(A \wedge dA + \frac{2}{3} A\wedge A\wedge A\right),
\end{equation} 
and $S_\inst$ is the contribution from worldsheet instantons with boundaries on $L$.
If $L$ is non-compact, then we should consider it as having a boundary $\partial L$ at infinity; the path integral on $L$ then gives a wave function in the Hilbert space of
Chern-Simons on the boundary.  The case of interest for the rest of this paper 
is $L \simeq \R^2 \times S^1$, which has
$\partial L = T^2$; from now on we specialize to that case, although the discussion
could be made more general. 

To find which state the topological open string theory picks, we need
to recall some facts about canonical quantization of the $U(M)$ Chern-Simons 
theory on $T^2\times \R$, viewing $\R$ as the ``time'' direction. 
We will be brief here; see e.g. \cite{Elitzur:1989nr} for more details. 
Integrating over the time component of the gauge field
localizes the path integral to flat connections on $T^2$:
\begin{equation} \label{cs-action-reduced}
\int \cD A' \;\delta(F') \;\exp \left(\frac{2\pi i}{k}\int_{T^2\times R} \Tr A' \,\partial_t \,A' dt\right).
\end{equation} 
Above $A'$ is a connection on $T^2$, which we can write (up to conjugation) as
\begin{equation}
A' = u\,d\theta_u + v\,d\theta_v,
\end{equation} 
where $u$ and $v$ are the components of $A'$ along two linearly independent
cycles of $T^2$, with intersection number $1$. 
From the action \eqref{cs-action-reduced} we see that $u$ and $v$ are
conjugate variables:  upon quantization we thus expect
\begin{equation}
[u,v] = i g_\Top,
\end{equation} 
where $g_\Top = \frac{2 \pi}{k+M}$.  The familiar shift of $k$ by $M$ 
can be seen by carefully integrating over massive modes \cite{Elitzur:1989nr}.

Since $u$ and $v$ are conjugate variables, in computing the Chern-Simons path integral
on a manifold with $T^2$ boundary, we should fix either $u$ or $v$ on the boundary, but not both, and the wave function will depend on the variable we have chosen to fix.  More 
generally, we could consider a mixed boundary condition where we fix
$v+{\overline \tau} u$ where ${\overline \tau}$ is some parameter (the motivation
for this notation will become clear later).

Note that in the present context $L \simeq \R^2 \times S^1$ is a solid torus,
so there is a unique $1$-cycle $\eta \in H_1(T^2,\Z)$ which collapses in the interior of $L$.
There is thus a canonical choice of polarization for the wave function; namely, one can
express it in terms of the holonomy around $\eta$,
which we call $v$.  In the next subsection we will relate this choice to the background
dependence (``holomorphic anomaly'') of the open topological string.
We could have tried to choose the ``cycle that survives'' in the interior of $L$ (corresponding
to the holonomy $u$), but this is ambiguous up to the shift $u \mapsto u + nv$.   This
ambiguity will be related to the framing ambiguity of the open topological string.

It can be shown \cite{Elitzur:1989nr, Aganagic:2002wv} that
the Chern-Simons path integral on the solid torus, without any
insertions and with $u$ fixed on the boundary, is given simply by
\begin{equation}
\psi_\Top^\open(u) = \langle L | u \rangle = 1.
\end{equation} 
In the present context, the Chern-Simons action is deformed by 
worldsheet instantons wrapping holomorphic curves with boundaries on $L$
\cite{Witten:1992fb}.  Their contribution to $S$ is
given by the free energy of the gas of topological open strings: 
\begin{equation}
S_\inst(u) = i F^{\open}_{\Top}(u).
\end{equation}
We now want to compute the path integral 
on $L$ with the operator insertion
\begin{equation}
\exp S_\inst(u). 
\end{equation} 
Since we are we are working in the basis
of eigenstates of $u$, the insertion just acts by multiplication:
\begin{equation}
\psi_\Top^\open(u) = 
\langle L |e^{S_\inst(u)}|u\rangle =e^{iF^{\open}_{\Top}(u)} \IP{L|u} = e^{iF^{\open}_{\Top}(u)}.
\end{equation} 
So we have identified the topological string partition function $e^{iF^{\open}_{\Top}(u)}$
with a wave function.

Although $v$ is the canonical choice,
we will sometimes find it natural to write the wave function in terms of one of the holonomies
$u+nv$ instead.  The relation between different choices of variable in which to write the wave function is
given by a Fourier transform:  for example, to transform from $u$ to $v$, one has
\begin{equation} \label{fourier}
\psi_\Top^\open(v) = \int d_H u \, e^{\frac{i}{g_{\Top}} \Tr uv} \, \psi_\Top^\open(u),
\end{equation}
where ${d_H}u$ is the measure induced from the Haar measure on $U(M)$.

The freedom to choose a variable is crucial because there are some cases in which the Lagrangian cycle 
$L$ can make a ``flop 
transition.''  From the perspective of the boundary $\partial L = T^2$ nothing special happens at the 
transition, but in the interior of $L$ the topology changes and in particular the cycle that collapses 
in the interior is different after the transition.  An example of this phenomenon can be seen when 
$X$ is a toric Calabi-Yau manifold.  Moreover, in that case one can use the
mirror B model to see that worldsheet instanton corrections eliminate the sharp
transition:  the different phases are smoothly connected.  
Thus, in the B model language there is a continuous change of variables
which takes us from one choice of holonomy to another.  This is
related to the background dependence of open topological string amplitudes,
to which we now turn.

\subsection*{Background dependence for the open topological string}

In this section we take a brief detour to explain the background dependence
of the open string topological string.\footnote{This was the original motivation for the present paper!}
It was conjectured in \cite{top-integ} that the open topological
string partition function depends on a choice of ``background'' moduli, or
equivalently, depends on the antiholomorphic coordinates of the moduli as well
as the holomorphic ones.
This conjecture was advanced in order to explain the fact that the open
topological string behaves like a wave function, by analogy to what is
known for the closed string case \cite{Witten:1993ed}.  
In the case considered
in \cite{top-integ}, the geometry of the Calabi-Yau is given (in the mirror B model)
by a hypersurface in $\C^4$,
\begin{equation}
F(u,v)-xy=0,
\end{equation} 
and the mirror of the Lagrangian brane is a brane on a holomorphic curve, 
specified by the condition $x=0$ together with fixed choices of $u$, $v$ 
satisfying $F(u,v)=0$.  As noted in \cite{Aganagic:2002wv,top-integ} the geometry with this
D-brane included can be viewed as a special (degenerate) limit of a closed string geometry,
with the D-brane serving as a source for the holomorphic 3-form; this source
changes the usual equation $d \Omega = 0$ to
\begin{equation}
d\Omega = g_\Top \delta (D),
\end{equation} 
where $\delta (D)$ denotes a delta function at the locus of the D-brane. 
We have already used this correspondence in Section \ref{sec-closed-revisited},
where we discussed how the ``ghost branes'' can be viewed as closed
string moduli.  Similarly, we can use it to interpret the holomorphic
anomaly of closed strings as inducing a holomorphic anomaly (or equivalently
a background dependence) for the open string partition function.   Here we
view the modulus of the open string, given by the choice of 
$(u,v)$ on the surface $F(u,v)=0$, as a closed string modulus.  In fact, borrowing the
closed string technology for background dependence developed in \cite{Witten:1993ed,Dijkgraaf:2002ac}
we immediately deduce that for a given background $(u_0,v_0)$ the natural
variable for the open string wave function is
\begin{equation} \label{hol-anomaly-good-var}
v+{\overline \tau} u,
\end{equation} 
where
\begin{equation}
{ \tau}={ {-\frac{\partial v}{\partial u}}} \Big\vert_{(u_0, v_0)}.
\end{equation}
Here we are considering $v$ as a function of $u$ through the implicit relation $F(u,v)=0$, so
$\tau$ is the slope of the tangent plane to the
Riemann surface at $(u_0,v_0)$.  Note that $\tau =\partial^2 F / \partial u^2$.  

The form \eqref{hol-anomaly-good-var} 
of the natural variable can be connected to our earlier discussion of the wave function
nature of the Chern-Simons theory embedded in the open string; there too
we claimed that there is a natural variable for the wave function, namely
the holonomy around the cycle of $T^2$ which shrinks in the interior of the
solid torus.  In that classical picture (which neglects the effect of worldsheet instantons)
the holonomy around the vanishing cycle is simply $v$; and choosing the background point 
near an asymptotic infinity of the quantum moduli space, where the classical picture
becomes exact, one indeed gets $\tau \to 0$, so $v + \overline{\tau} u \to v$.
More invariantly, the value of $\tau$ near an asymptotic infinity of the B model Riemann surface 
approaches the slope of the corresponding line in the A model toric diagram, and this slope indeed determines
the collapsing cycle of the toric fiber.  

Note that in order to go off the real locus $\tau = \overline{\tau}$ 
we need to recall that the Chern-Simons holonomies
are complexified in the context of topological strings (to include the
moduli which move the brane); in the geometric motivation
we gave before we had essentially turned those off.  It would be interesting
to understand this relation off the real locus.

\section{The open string conjecture} \label{sec-open}

As we reviewed in Section \ref{sec-closed}, the closed topological string wave function
on a Calabi-Yau space $X$ is believed
to compute the large-charge asymptotics of an index which counts
BPS states in four dimensions, and this index has an
interpretation as the Wigner function of $\psi_\Top$.  On the other hand, we just
saw in Section \ref{sec-qm-open} that the open topological string partition function $\psi_\Top^\open$
with non-compact D-branes is also naturally considered as a wave function.  So we could construct
a Wigner function from this wave function, and then a 
natural question is whether this Wigner function also has an interpretation as counting
BPS states.  We will argue that it does.

We embed the open topological string
in the superstring in a familiar way \cite{Ooguri:1999bv}.  
Namely, consider D4-branes wrapping a special Lagrangian
cycle $L \subset X$.  Then there are open D2-branes
ending on these D4-branes.  These
give rise to BPS particles in the 2-dimensional supersymmetric
gauge theory on the non-compact directions of the D4-branes; we will interpret 
the charge $\cQ_e$ as counting these BPS particles.  The gauge theory in question
also supports BPS domain walls; we will interpret $\cQ_m$ as measuring the domain wall 
charge.

Altogether then, we will conjecture below that the
open topological string, on a Calabi-Yau space $X$ with Lagrangian branes
included, computes the large-charge asymptotics of an index which counts open 
D2-branes, and their domain wall counterparts, bound to any number of closed
D0, D2, D4 and D6-branes.  Furthermore, we will describe one context in which
some aspects of this proposal can be checked.

\subsection*{Calabi-Yau spaces with branes and BPS particles}

Consider a Calabi-Yau manifold $X$ containing a special Lagrangian
3-cycle $L$. We consider the Type IIA superstring on $X \times
\R^{3,1}$, with $M$ D4-branes on $L \times \R^{1,1}$, which we will
call the ``background branes.''  For simplicity, we assume $L$ has the
topology
\begin{equation}
L \simeq \R^2 \times S^1.
\end{equation}

The dimensionally reduced theory on the $\R^{1,1}$ part of the background 
branes is a $(2,2)$ supersymmetric gauge theory.
Its field content can be understood as follows 
\cite{Ooguri:1999bv}.  Since $b_1(L)=1$ it follows \cite{MR1664890}
that $L$ has one real modulus $r$; this modulus pairs up with the Wilson line of the 
worldvolume gauge field $\oint A$ to give a complex field 
\begin{equation}
u=r + i \oint A.
\end{equation}  
One also gets a gauge field in $\R^{1,1}$ by
integrating the world-volume two-form $B$ (which is the magnetic dual
to the gauge field $A$ on the D4-brane, defined by $d\!*\!A = dB$) over the
$S^1$ of $L$.  Since there are $M$ D4-branes, the theory has (at least) a magnetic $U(1)^M$
gauge symmetry.  The field $u$ should be viewed as the lowest
component of a twisted chiral multiplet, whose top component is the
field strength of the magnetic gauge field in two dimensions.

There is an obvious way of getting BPS particles in this theory.
Suppose for a moment that $M=1$ (a single Lagrangian brane.)
Let $\gamma \in H_1(L,\Z)$ denote the homology class of the $S^1$ in $L$.
Since the Calabi-Yau has no non-contractible 1-cycles, this $\gamma$ is a
boundary in $X$; so there exists some $D$ with
\begin{equation}
[ \partial D ] = \gamma.
\end{equation} 
Open D2-branes wrapped on $D$ give rise to particles charged under the $U(1)$ gauge field
of the 2-dimensional theory; if $D$ is a holomorphic disc, then these particles are BPS.

\subsection*{The conjecture}

Now, to motivate our conjecture, recall from Section \ref{sec-closed} 
that in the closed string case (without the background branes) we have the relation
\begin{equation} \label{bh-partition-function-redux}
Z_\BH(Q_6, Q_4, \varphi_2, \varphi_0) = \sum_{Q_0, Q_2} \Omega_{Q_6, Q_4, Q_2, Q_0} e^{-Q_2 \varphi_2 - Q_0 \varphi_0}=\abs{\psi_\Top(g_\Top, t)}^2,
\end{equation}
where $\varphi_2 = \im 2 \pi t /g_\Top$ and $\varphi_0 = \im 4\pi^2 / g_\Top$ as given in 
\eqref{osv-coupling}, \eqref{osv-kahler}.
We wish to generalize this conjecture to the open topological string.
What is the appropriate ensemble to consider?
Since the closed D2-branes are ``light electric states'' in the closed
string ensemble, which we sum over with chemical potentials, 
it is natural to try treating the open D2-branes in the same way.  
Thus, in formulating our conjecture we consider these BPS states 
as ``electric charges,'' and sum over them with a chemical potential $\varphi_e^{\open}$.
We also expect to have a ``magnetic charge,'' which we
fix to the some value $\cQ^\open_m$; we will discuss these
charges further below.\footnote{The terminology ``electric'' and ``magnetic'' here is chosen
by analogy to the closed string case.  The charges we are discussing here are both associated 
to point particles, which are not electric-magnetic duals in the theory on $\R^{1,1}$.}
The partition function of the ensemble thus obtained is a simple generalization of \eqref{bh-partition-function-redux},
\begin{equation} \label{bh-partition-function-open}
Z^\open_\BPS(Q_6, Q_4, \cQ^\open_m, \varphi_2, \varphi_0, \varphi_e^\open) = \sum_{Q_0, Q_2, \cQ^\open_e} \Omega_{Q_6, Q_4, Q_2, Q_0, \cQ^\open_e, \cQ^\open_m} e^{-Q_2 \varphi_2 - Q_0 \varphi_0 - \cQ^\open_e \varphi^\open_e}.
\end{equation}
We conjecture that the relation of $Z^\open_\BPS$ to the topological string is a direct
generalization of \eqref{bh-partition-function-redux},
\begin{equation} \label{osv-conjecture-open}
Z^\open_\BPS(Q_6, Q_4, \cQ^\open_m, \varphi_2, \varphi_0, \varphi^\open_e) = 
\abs{\psi_\Top^\open(g_\Top, t, u)}^2,
\end{equation} 
where $\psi_\Top^\open$ is the topological A model partition function on $X$, including
open strings ending on $M$ D-branes on $L$ as well as closed strings.

In this conjecture the closed string moduli $g_\Top, t$ are determined by the attractor mechanism as before.  What about the open string modulus $u$?  
The formula $\varphi_2 = \im 2 \pi t / g_\Top$ for
the closed D2-brane chemical potential suggests that the open D2-brane chemical potential
should be related to $u$ by
\begin{equation}
\varphi_e^{\open} = \im 2 \pi u / g_\Top.
\end{equation}
We will verify this identification of $\im u$ in an explicit example below.
The real part of $u$ should be fixed by the charge $\cQ^\open_m$, as we now discuss.

\subsection*{Adding magnetic charges}

What is the spacetime meaning of the ``magnetic'' charge $\cQ^\open_m$, and 
its relation to the real part of the modulus $u$?  We 
can make a plausible guess by exploiting the symmetry between $u$ and its conjugate $v$.
Namely, as noted in Section \ref{sec-qm-open}, it is possible for $L$ to undergo a flop
transition to a new phase parameterized by a different parameter $v$ (representing the holonomy
of the gauge field around a new $S^1$ which was contractible in the old phase).  The two phases are
smoothly connected in the quantum topological string theory and also in the physical
one, but they correspond to different classical descriptions of the physics.  The most economical 
assumption would then be that the excitations which we are calling ``electric'' in one description 
are the same as the ones which we are calling ``magnetic'' in the other.  In this section
we explore the consequences of this assumption (without being too careful about the factors
of $i$ which appear.)  We discuss only the open string sector, suppressing the closed strings,
and drop the label ``open'' from our notation for simplicity.

First, we can write down the precise form of $u$, using 
the fact that $\psi_\Top(v)$ is related to $\psi_\Top(u)$ by the 
Fourier transform \eqref{fourier}, or equivalently
\begin{equation} \label{commutation-1}
[u, v] = i g_\Top.
\end{equation}
The dictionary between our statistical ensemble and the quantum-mechanical
picture requires the relations
\begin{equation} \label{commutation-2}
[\cQ_e,\varphi_e] = 1 = [\cQ_m,\varphi_m],
\end{equation} 
since we cannot fix the charges and the chemical potentials at the same time.  On the other hand,
we can fix the charges simultaneously, so
\begin{equation} \label{commutation-3}
[\cQ_e,\cQ_m] = 0 = [\varphi_e,\varphi_m].
\end{equation}
The consistency of \eqref{commutation-1}, \eqref{commutation-2}, \eqref{commutation-3} 
with $\im u = g_\Top \varphi_e / 2 \pi$ then requires
\begin{equation}\label{attractor}
\re u = \pi \cQ_m, \quad \quad
\re v = \pi \cQ_e.
\end{equation}
The equation \eqref{attractor} completes our conjecture \eqref{osv-conjecture-open}, except
that we have not been precise about how to fix the zero of $\re u$ or $\re v$.  We do not have a general
proposal for how this should be done, although we will see how it works in an example below.

Note that the expectation value of $v$ in the state corresponding to the 
open string wave function $\psi_\Top(u)= \exp(i F_{\Top}(u))$
is given by
\begin{equation}
v = g_s \partial_u F_{\Top}(u)
\end{equation} 
(the semi-classical version of this equation was discovered in \cite{Aganagic:2001nx}).
This is precisely analogous to the special geometry relations of the closed string.
In this sense \eqref{attractor} seems to describe an open string
analogue of the attractor mechanism that fixes the moduli to values
determined by charges of BPS states.  It would be interesting to study
this attractor mechanism directly in the physical theory.

Our identification of the parameters leads to two formulas for the Wigner function, i.e.
the degeneracies of BPS states,
\begin{align}
\Omega_{\cQ_e,\cQ_m} &= \int d\varphi_e \; e^{- \cQ_e \varphi_e}\;
\psi(u = \frac{i g_\Top}{2\pi} \varphi_e + \pi \cQ_m)\overline{\psi(u = \frac{i g_\Top}{2\pi} \varphi_e+\pi \cQ_m)},\\
&= \int d\varphi_m \; e^{- \cQ_m \varphi_m}\;
\psi(v = \frac{i g_\Top}{2\pi} \varphi_m + \pi \cQ_e)\overline{\psi(v = \frac{i g_\Top}{2\pi} \varphi_m+\pi \cQ_m)}.
\end{align}
(The arguments we gave above about commutation relations are equivalent to the statement that these 
two formulas are indeed related by Fourier transforming $\psi(u) \leftrightarrow \psi(v)$.)
Put another way, $\psi(u)$ and $\psi(v)$ sum over conjugate ensembles,
\begin{align}
\abs{\psi(u = \frac{i g_\Top}{2\pi} \varphi_e + \pi \cQ_m)}^2 &= \sum_{\cQ_e} \Omega_{\cQ_e,\cQ_m} e^{- \cQ_e \varphi_e}, \label{ensemble-u} \\
\abs{\psi(v = \frac{i g_\Top}{2\pi} \varphi_m + \pi \cQ_e)}^2 &= \sum_{\cQ_m} \Omega_{\cQ_e,\cQ_m} e^{- \cQ_m \varphi_m}.
\end{align} 
In the above we implicitly chose some framing for the open string wave function $\psi(u)$, and one
could ask what is the meaning of changing the framing.  As discussed in \cite{top-integ},
the effect of shifting the framing by $k$ units is $\psi^{(k)}(u) = e^{- i k g_\Top \partial^2_u} \psi(u)$.
From this and \eqref{ensemble-u} 
it follows that $\psi^{(k)}$ sums over an ensemble in which we have a chemical potential for
dyons of charge $(1,k)$:
\begin{equation}
\abs{\psi^{(k)}(u = \frac{i g_\Top}{2\pi} \varphi_e + \pi \cQ_m)}^2 = \sum_{\cQ_e} \Omega_{\cQ_e,\cQ_m + k\cQ_e} e^{- \cQ_e \varphi_e}.
\end{equation}

So far we have discussed the magnetic charge $\cQ_m$ abstractly in terms of its relation 
to the real part of the topological string modulus, but our assumption also 
leads to a natural description of the meaning of the magnetic charges in the physical theory.
To understand this, note first that turning on electric charge
$\cQ_e$, arising from open D2-branes ending on $\gamma \subset L$, can be 
equivalently described as turning on magnetic flux on the background D4-brane. 
This is because the D2-brane ending on $L$ looks like a monopole string from the point 
of view of the gauge theory on the D4-brane.  So, letting $D$ denote any 2-cycle in 
$L$ dual to $\gamma$ (see Figure \ref{fig-dual-cycle-outline-fonts-small}), we have
\fig{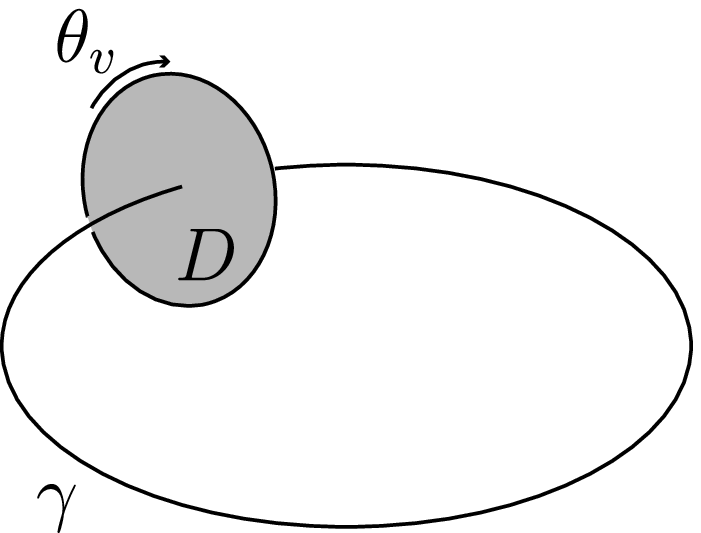}{The 1-cycle $\gamma$ and its dual cycle $D$ inside $L$.}{height=1.5in}
\begin{equation} \label{flux}
\int_{\R \times D} dF = 2\pi \cQ_e,
\end{equation} 
where $\R$ denotes the spatial $x$-direction in $\R^{1,1}$.
In particular, we could choose $D$ to be the disc obtained by filling in 
the 1-cycle $S^1$ corresponding to $v$, which opens up after the flop transition.  
Then \eqref{flux} is equivalent to
\begin{equation}
\int_{\R \times \partial D} F_{x \theta_v}\,dx\,d \theta_v  = 2\pi \cQ_e. 
\end{equation} 
Alternatively, as $F_{x \theta_v} = \partial_x A_{\theta_v}$ and
\begin{equation}
\oint_{\partial D} A_{\theta_v}\,d\theta_v = {\im} v,
\end{equation} 
we see that as we cross the D2-branes in the $x$ direction $v$ jumps by $2\pi i \cQ_e$.
Since exchanging electric and magnetic charges corresponds to
exchanging $u$ and $v$, it follows that turning on $\cQ_m$
units of magnetic charge corresponds to having a domain wall where
$u$ jumps by $2\pi i \cQ_m$ in going from $x=-\infty$ to $x=+\infty$. 
Hence these domain walls are the magnetic charges we were seeking.

\subsection*{Multiple Lagrangian branes}

In the above discussion we have been assuming that we have a single background D4-brane.
Let us now return to more general case $M\geq 1$.  In this case ${\cQ}^\open_{e,m}$ label representations of $U(M)$.\footnote{The gauge theory has at least a $U(1)^M$
symmetry, and since the degeneracies are symmetric under the symmetric group $S_M$, 
we can organize them into characters of representations $\cQ^\open_e$ of $U(M)$ 
(possibly with negative multiplicities.)}
By a straightforward generalization of the arguments given above, 
we see that the attractor values of the eigenvalues of
$u$ and $v$ are (generalizing \eqref{attractor})
\begin{equation}
\re u_i = \pi ({{\hat\cQ}^\open}_{m})_{i}, \quad \quad
\re v_j = \pi ({{\hat\cQ}^\open}_{e})_{j}.
\end{equation}
Here ${{\hat\cQ}^\open}_{m,e}$ denote the highest weight vectors
of the corresponding representations, shifted by the Weyl vector $\rho$ (see Appendix \ref{app-group-theory}).  The rest of the discussion generalizes similarly.

\section{A solvable example} \label{sec-2dym-open}

After these general considerations we now return to the example we described in Section \ref{sec-closed},
where $X$ is a rank 2 holomorphic vector bundle over $\Sigma$, and add background D4-branes on $L \times \R^{1,1}$ 
to the Type IIA theory.  In this section we want to argue that one can use 2-dimensional 
Yang-Mills theory to compute $Z^\open_\BPS$, generalizing the discussion
of Section \ref{sec-closed}.  Our arguments will be heuristic, but they lead to a definite
prescription which is natural and fits in well with our conjectures.

How is the discussion of Section \ref{sec-closed} 
modified by the introduction of the background branes?
The $L$ we will consider meet $\cD$ along a circle, which we call
$\gamma$.  Hence
in the gauge theory on $\cD$ 
there will be extra massless string states localized along $\gamma$, 
in the bifundamental of $U(M) \times U(N)$.  By condensing
these string states (going out along a Higgs branch), i.e. turning on a vacuum expectation value of the form
\begin{equation}
\begin{pmatrix}
1 & & \\
& \ddots & \\
& & 1 \\
0 & \cdots & 0 \\
\vdots & & \vdots \\
0 & \cdots & 0
\end{pmatrix}
\end{equation}
one can break the gauge symmetry along $\gamma$ to $U(M) \times U(N-M)$, where the surviving $U(M)$ is the 
diagonal in $[U(M) \times U(M)] \times U(N-M)$.\footnote{We are considering only the case $M<N$; ultimately
we will be interested in taking $N$ large while $M$ stays finite.}  
We conjecture that from the point of view 
of the gauge theory on $\cD$, the only effect of the interaction with the background branes
comes from the fact that the $U(M)$ part of the gauge field along $\gamma$ is identified with the $U(M)$ gauge field
on the background branes, via this Higgsing to the diagonal.  We can account for this by inserting a $\delta$-function
in the theory on $\cD$, which freezes $M$ of the eigenvalues of the holonomy
$e^{i \oint_\gamma \cA}$, identifying them with the holonomy on the background branes, which we call $e^{i \phi}$.  The Weyl invariant way to write this delta function is
\begin{equation} \label{delta-function-4d}
\delta_M \left(e^{i \oint_\gamma \cA}, e^{i \phi} \right) = D(\oint_\gamma \cA)^{-1} \sum_{\sigma \in S_N} (-)^\sigma \prod_{j=1}^M \delta \left((e^{i \oint_\gamma \cA})_{\sigma(j)}, e^{i\phi_j} \right),
\end{equation}
where $D$ denotes the Vandermonde determinant \eqref{vandermonde}.

Let us write $Z_\YM^\open(\varphi_0, \varphi_2, \phi)$ for the partition function with this operator inserted (here ``open'' refers to the fact that it is related to the open topological string.)  
This partition function sums
over the open D2-branes which end on the Lagrangian branes, as well as over the
D0 and D2-brane charges which one had without the Lagrangian branes; 
so altogether we should have
\begin{equation}
Z_\YM^\open = Z_\BPS^\open.
\end{equation}
In this ensemble the chemical potential $\varphi_e^\open$ for the open D2-branes should roughly
be their mass.  This mass is given by the area of the disc on which they are wrapped, which is related by supersymmetry to the Wilson line on the background branes; with this as motivation we write
\begin{equation}
\varphi_e^\open = 2 \pi \phi / g_s.
\end{equation}

To compute $Z_\YM^\open$ it is convenient to reduce from the twisted $\N=4$ theory on $\cD$ to 
a $q$-deformed Yang-Mills theory on $\Sigma$, as was done in \cite{Vafa:2004qa,Aganagic:2004js}.
How does the operator insertion $\delta_M \left(e^{i \oint_\gamma \cA}, e^{i \phi} \right)$ translate to the reduced theory?  
There are two cases to consider:
\begin{enumerate}
\item{$\gamma$ lies in the fiber of $\LL_{-p}$ over a point $P \in \Sigma$.}
\item{$\gamma$ lies on the Riemann surface $\Sigma$.}
\end{enumerate}
In either case these Lagrangian branes can be locally modelled by the
ones studied in \cite{Aganagic:2000gs, Aganagic:2001nx}. 
In case 1, where $\gamma$ is in the fiber over $P$, the situation is basically straightforward:  as explained
in \cite{Vafa:2004qa}, the flux $\oint_\gamma \cA$ shows up in the $q$-deformed Yang-Mills
theory on $\Sigma$ as a field $\Phi$.  The operator we have to insert in the $q$-deformed theory
is therefore
\begin{equation} \label{delta-function-fiber}
\delta_M(e^{i \Phi(P)}, e^{i \phi}).
\end{equation}
The path integral gets localized on configurations where $\Phi$ is locally constant, so when there
are no other operator insertions we can drop the $P$ and write $\delta_M(e^{i \Phi}, e^{i \phi})$.

In case 2 the situation is a bit trickier, because of a subtlety which also appeared in \cite{Vafa:2004qa}:
namely, in performing the reduction one has to choose $p$ points $P_i$ on $\Sigma$, and at each such point
one gets an operator corresponding to one unit of area in the Yang-Mills theory.  The operator
$\delta_M \left(e^{i \oint_\gamma \cA}, e^{i \phi} \right)$ reduces to
\begin{equation} \label{delta-function-base}
\delta_M\left(e^{i \oint_{\gamma} A}, e^{i \phi}\right)
\end{equation}
in two dimensions, but we have to specify how many
of the $p$ points go on each side of $\gamma$.  Therefore there is a $\Z$-valued ambiguity
in defining which operator we insert in the physical theory, parameterized by a choice of $p_1$ and $p_2$ with $p_1 + p_2 = p$.  See Figure \ref{fig-ambiguity-small}.  This ambiguity should be understood
as related to infrared regularization arising from the non-compactness of the situation; in
the connection to the open topological string below, we will see that it is identified with the framing ambiguity.
\fig{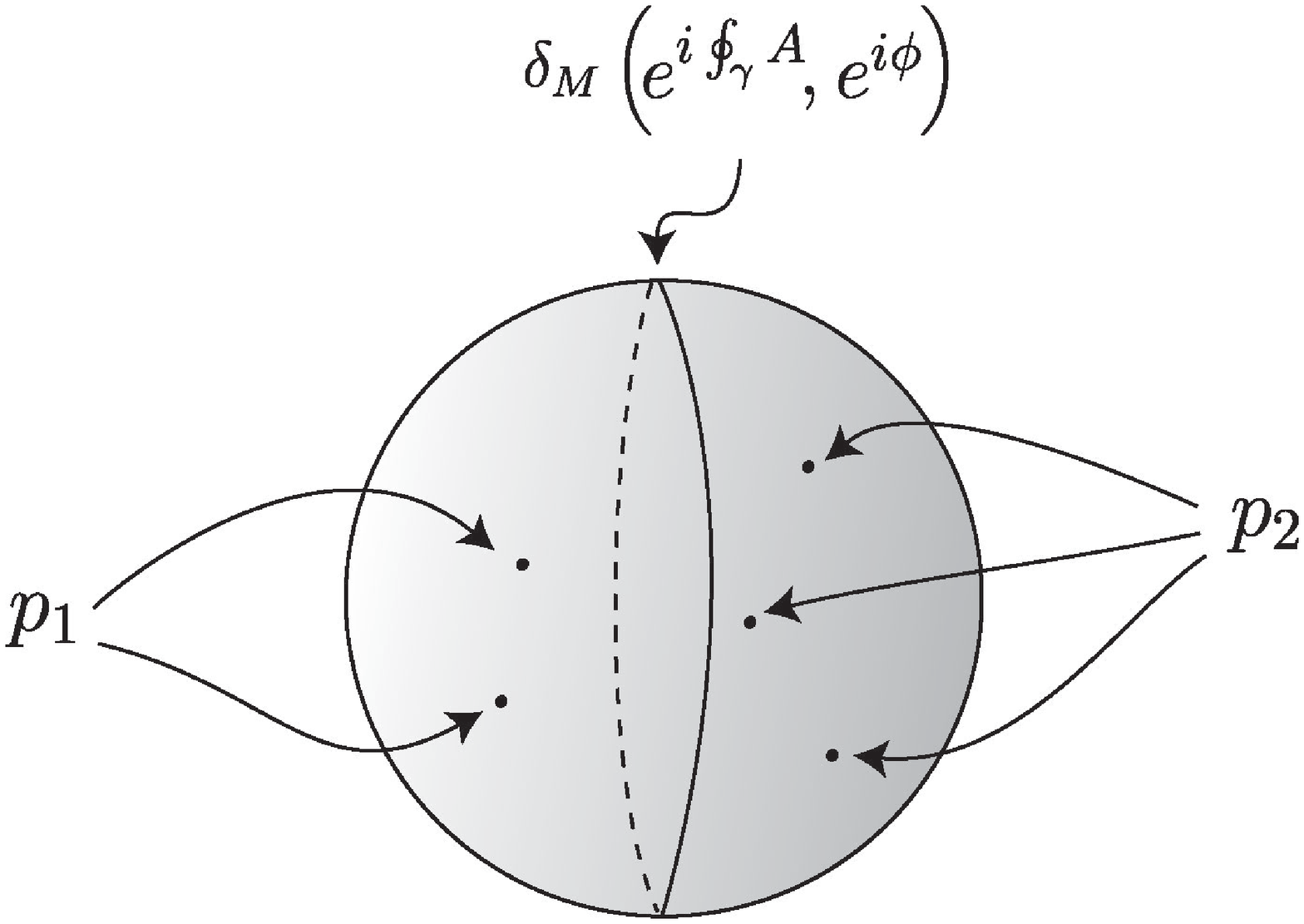}{The operator $\delta_M\left(e^{i \oint_{\gamma} A}, e^{i \phi}\right)$ cuts
$\Sigma$ into two pieces.}{height=2.4in}

\subsection*{Specializing to genus zero}

Next we will investigate in detail the case when $\Sigma$ has genus zero.
So we specialize to Type IIA on $X \times \R^{3,1}$, where
\begin{equation} \label{local-cy-sphere}
X = \OO(-p) \oplus \OO(p-2) \to \C\PP^1,
\end{equation}
with background D4-branes added on $L \times \R^{1,1}$.  As we just explained, we
can compute a mixed ensemble partition function $Z^\open_\BPS$ for this system by inserting an appropriate
operator into the $q$-deformed Yang-Mills theory on an $S^2$ of area $p$.  
The parameters of the Yang-Mills theory are as in the closed case,
\begin{equation} \label{2dym-params-open}
\theta_{\YM} = \theta, \quad g^2_{\YM} = g_s, \quad q = e^{-g_s}.
\end{equation} 

We will show that for all $p_1$, $p_2$ we indeed have 
$Z_\YM^\open = \abs{\psi_\Top^\open}^2 + \OO(e^{-N})$.  We will also show that the identification of $Z_\YM^\open$ with $Z_\BPS^\open$ is consistent; 
namely, $Z_\BPS^\open$ should have an expansion where $\phi$ appears 
only in the form $e^{- 2 \pi \phi / g_s}$, and we will
verify that $Z_\YM^\open$ indeed has such an expansion at least in the special case $p_1 = p_2 = 1$.
These two results together give evidence for our conjecture \eqref{osv-conjecture-open}.

\subsection*{Large $N$ factorization on $\OO(-p) \oplus \OO(p-2) \to \C\PP^1$}

We want to establish that 
\begin{equation}
Z_\YM^\open = \abs{\psi_\Top^\open}^2 + \OO(e^{-N}).
\end{equation}  
We compute $Z_\YM^\open$ using the gluing procedure described in Appendix \ref{app-qym}:  namely, we construct the sphere by gluing two discs together with the operator $\delta_M(e^{i \oint A}, e^{i \phi})$ in the middle.  We use the fact that the Hilbert space of the 2-dimensional Yang-Mills theory is factorized at large $N$, $\HH \simeq \HH_+ \otimes \HH_-$, and furthermore
each component of the gluing procedure can be written in 
a factorized form.  This factorization is described in detail in Appendix \ref{app-factorization}; the computation of $Z_\YM^\open$ we give below basically consists of fetching various results from that appendix and putting them together.  We then compare this with the known form of the topological string amplitude and find the desired factorization; the final result is given in \eqref{2dym-factorization-open}.

\subsubsection*{Branes in the base}

Let us first discuss case 2, where to compute $Z_\YM^\open$ we have to insert a Wilson line
freezing operator $\delta_M(e^{i \oint A}, e^{i\phi})$.  This operator cuts the sphere into two pieces, with discrete areas $p_1$, $p_2$ such that $p_1 + p_2 = p$.  The gluing computation of $Z_\YM^\open$ involves a zero area disc, an annulus of area $p_1$, the operator $\delta_M(e^{i \oint A}, e^{i\phi})$, an annulus of area $p_2$, and another zero area disc:
\begin{equation} \label{gluing-base}
Z^\open_{\YM}(N, g_s, \theta, \phi) = \IP{\Psi_0 | A_{p_1} \delta_M(e^{i \oint A}, e^{i\phi}) A_{p_2} | \Psi_0}. 
\end{equation}
Each of these pieces has been written in the factorized basis for $\HH$ in Appendix \ref{app-factorization}:  the disc
is given in \eqref{disc-factorized}, the annulus in  \eqref{annulus-factorized}, and the Wilson line freezing operator
in \eqref{wilson-line-freeze-factorized}.
Plugging in these factorizations 
and doing a little rearranging, we obtain the factorized form of $Z_\YM^\open$, 
schematically $Z^\open_\YM = Z_+ Z_-$, or more precisely (writing $q = e^{- g_s}$)
\begin{multline} \label{ym-sphere-wilson-line-freeze-factorized}
Z^\open_{\YM}(N, g_s, \theta, \phi) = Z_{\YM}^0(N, g_s, \theta, \phi) M(q)^2 \eta(q)^{2N} \times \\
\sum_{l \in \Z, R'_1, R'_2} (-)^{\abs{R'_1}+\abs{R'_2}} q^{\half(p_1+p_2)l^2} e^{iNlp\theta} 
Z^{R'_1, R'_2,l}_+ Z^{R'_1, R'_2,l}_- + \OO(e^{-N}),
\end{multline}
where
\begin{multline} \label{chiral-block-sphere-wilson-line-freeze}
Z^{R'_1, R'_2,l}_+(N, g_s, \theta, \phi) = q^{\half N (\abs{R'_1}+\abs{R'_2})} \times \\ 
\sum_{R_{1+}, R_{2+}, A_+} q^{\half p_1 \kappa_{R_{1+}} + \half p_2 \kappa_{R_{2+}} + (\half N(p_1-1)+lp_1) \abs{R_{1+}}+(\half N(p_2-1)+lp_2) \abs{R_{2+}}}  \times \\
C_{0 R'_1 R_{1+}} C_{R'^t_2 R_{2+} 0} \schur_{R_{1+} / A_+} (e^{-i\phi}) \schur_{R_{2+} / A_+} (e^{i\phi}) e^{i \theta (p_1 \abs{R_{1+}} + p_2 \abs{R_{2+}} ) },
\end{multline}
and similarly
\begin{multline} \label{antichiral-block-sphere-wilson-line-freeze}
Z^{R'_1, R'_2,l}_-(N, g_s, \theta, \phi) = q^{\half N (\abs{R'_1}+\abs{R'_2})} \times \\
\sum_{R_{1-}, R_{2-}, A_-} q^{\half p_1 \kappa_{R_{1-}} + \half p_2 \kappa_{R_{2-}} + (\half N(p_1-1)-lp_1) \abs{R_{1-}}+(\half N(p_2-1)-lp_2) \abs{R_{2-}}}  \times \\
C_{0 R'^t_1 R_{1-}} C_{R'_2 R_{2-} 0} \schur_{R_{1-} / A_-} (e^{i\phi}) \schur_{R_{2-} / A_-} (e^{-i\phi}) e^{-i \theta (p_1 \abs{R_{1-}} + p_2 \abs{R_{2-}} ) }.
\end{multline}
The normalization factor $Z_{\YM}^0(N, g_s, \theta, \phi)$ will be fixed below.

Now we want to interpret the chiral blocks $Z^{R'_1, R'_2, l}_\pm(\phi)$ in terms of the topological string on $X$.
This $X$ can be represented torically by the picture in Figure \ref{fig-toric-internal}, on which we also indicate
the Lagrangian cycle $L$ supporting $M$ branes, one supporting a stack of infinitely many ghost branes, and
one with a stack of infinitely many ghost antibranes.
(See e.g. \cite{Aganagic:2000gs} for a review of the meaning of toric pictures such as this one.)

\begin{figure}[t]
\centering
\psset{unit=0.30}
\input{fig-toric-internal.tex}
\caption{The vertex representation of $X = \OO(-p) \oplus \OO(p-2) \to \C\PP^1$, 
with a stack of $M$ branes with
complexified holonomy $U = e^u$, a stack of infinitely many ghost branes with complexified holonomy $U'_1 = e^{u'_1}$,
and a stack of infinitely many ghost antibranes with complexified holonomy $U'_2 = e^{u'_2}$.}
\label{fig-toric-internal}
\end{figure}

The results of \cite{Aganagic:2003db} give the topological string amplitude on this geometry as\footnote{Using the result as it appears in \cite{Aganagic:2003db} one would actually get something slightly different from \eqref{top-sphere-internal-line}, namely, $R_2 / A$ would be replaced by $R_2^t / A^t$, and there would be an extra overall factor $(-)^{\abs{R_2} + \abs{A}}$.  This difference is due to a typo in \cite{Aganagic:2003db}.} (with $q = e^{-g_\Top}$)
\begin{multline} \label{top-sphere-internal-line}
\psi_{\Top}^{\ghost}(g_\Top,t,u,u') = \psi_\Top^0(g_\Top,t,u) \sum_{R_1, R_2, A, R'_1, R'_2} (-)^{\abs{R'_2}} \schur_{R'_1}(e^{u'_1}) \schur_{R'_2}(e^{u'_2}) \times \\
q^{\half p_1 \kappa_{R_1} + \half p_2 \kappa_{R_2}}  C_{0 R'_1 R_1} C_{R'^t_2 R_2 0} \schur_{R_1 / A}(e^{-u}) \schur_{R_2 / A}(e^{u}) (-)^{p_1\abs{R_1} + p_2 \abs{R_2}} e^{-t\abs{R_2}}, 
\end{multline}
with the choice of $p_1$ and $p_2$ (subject to the constraint $p_1 + p_2 = p$) 
related to the choice of framing on the Lagrangian branes.\footnote{Strictly speaking, \cite{Aganagic:2003db} considers the case $M = \infty$; but one can get finitely many branes by setting all but $M$ components of the $e^u$ and $e^{-u}$ appearing in \eqref{top-sphere-internal-line} to zero.}  Similarly, if one swaps the ghost branes for ghost antibranes, one gets
\begin{multline} \label{top-sphere-internal-line-antibranes}
\psi_{\Top}^{\antighost}(g_\Top,t,u,u') = \psi_\Top^0(g_\Top,t,u) \sum_{R_1, R_2, A, R'_1, R'_2} (-)^{\abs{R'_1}} \schur_{R'_1}(e^{u'_1}) \schur_{R'_2}(e^{u'_2}) \times \\
q^{\half p_1 \kappa_{R_1} + \half p_2 \kappa_{R_2}}  C_{0 R'^t_1 R_1} C_{R'_2 R_2 0} \schur_{R_1 / A}(e^{-u}) \schur_{R_2 / A}(e^{u}) (-)^{p_1\abs{R_1} + p_2 \abs{R_2}} e^{-t\abs{R_2}}.
\end{multline}

Now to relate the chiral blocks $Z_\pm$ which make up $Z_\YM$ to the topological string amplitudes, we define
\begin{align}
t &= \half N g_s(p - 1) - i p \hat\theta, \label{params-t1} \\
u &= \half N g_s(p_1 - 1) - i (p_1 \hat\theta - \phi),\\
u'_1 &= \half N g_s + i \phi'_1, \\
u'_2 &= \half N g_s + i \phi'_2, \\
g_\Top &= g_s.
\end{align} 
Here we introduced $\hat\theta = \theta + \pi$; this shift is meant to cancel the factor $(-)^{p \abs{R}}$ in \eqref{top-sphere-internal-line}.\footnote{The apparent asymmetry between $p_1$ and $p_2$ comes from the fact that we chose $u$ to represent the complexified area of the disc which ends on the Lagrangian branes from the left; the disc which ends on them from the right has area $t-u = \half Ng_s(p_2 - 1) - i 
(p_2 \hat{\theta} + \phi)$.}

The desired factorization is then basically straightforward to check.  One begins with 
\eqref{ym-sphere-wilson-line-freeze-factorized} which expresses $Z_\YM$ in terms of the
chiral blocks, then relates the chiral blocks to $\psi_\Top^\ghost$ and $\psi_\Top^\antighost$
with the above choice of parameters, and converts
the sums over $R'_1$, $R'_2$ into integrals over $\phi'_1$, $\phi'_2$ as discussed in
Section \ref{sec-closed-revisited}.  This essentially gives
\begin{multline} \label{2dym-factorization-open}
Z_{\YM}(N, g_s, \theta, \phi) = \sum_{l \in \Z} \int d_H \phi'_1 d_H \phi'_2 \times \\ 
\left(\psi_\Top^{\ghost} \left(g_s, t + l p g_s, u + l p_1 g_s, u' \right) \right) \overline{\left(\psi_\Top^{\antighost} \left(g_s, t - l p g_s, u - l p_1 g_s, u' \right) \right)}.
\end{multline}
In order to match the $l$-dependent terms in \eqref{ym-sphere-wilson-line-freeze-factorized}, 
though, one has to examine carefully the normalizations for the topological string and Yang-Mills 
amplitudes, as was done in \cite{Vafa:2004qa,Aganagic:2004js}.  For the topological string we write
\begin{equation} \label{top-normalization-sphere}
\psi_\Top^0(g_\Top, t) = M(q) \eta(q)^{2 t / (p-2)g_\Top} \exp \left( - \frac{1}{6 p(p-2) g_\Top^2} t^3 + \frac{p-2}{24p} t \right).
\end{equation}
The meaning of this normalization factor was discussed in \cite{Aganagic:2004js}.
For the Yang-Mills theory, we write
\begin{equation} \label{ym-normalization-sphere}
Z_{\YM}^0(N, g_s, \theta, \phi) = \exp \left(\frac{g_s (p-2)^2}{24 p} (N - N^3) + N \frac{\hat\theta^2 p}{2 g_s} \right).
\end{equation}
Then the two chiral normalization factors multiply together to give the Yang-Mills normalization, up to some crucial $l$-dependent corrections:
\begin{multline}
\psi^0_\Top(g_s, t + l p g_s, u + l p_1 g_s) \overline{\psi^0_\Top(g_s, t - lp g_s, u - l p_1 g_s)} = \\
Z^0_{\YM}(N,g_s,\theta,\phi) M(q)^2 \eta(q)^{2N} q^{\half p l^2} e^{i N l p \theta}.
\end{multline}
These terms match the $l$-dependent terms in \eqref{ym-sphere-wilson-line-freeze-factorized}; they
are exactly what is needed to make the factorization \eqref{2dym-factorization-open} work.
So we have completed the factorization in case 2, corresponding to D-branes which intersect the base $\C\PP^1$ in $X$.

\subsubsection*{Branes in the fiber}

We can also consider case 1, corresponding to D-branes which meet the fiber
of $\OO(-p) \to \C\PP^1$.  In this case, in the Yang-Mills theory we insert the dual 
Wilson line freezing operator $\delta_M(e^{i \Phi}, e^{i \phi})$ at a point of $\C\PP^1$.  Our discussion here will be 
more brief since the proof of the factorization runs along the same lines as case 2 above.

Again, we compute the Yang-Mills amplitude by gluing:  we have to glue a disc containing the operator $\delta_M(e^{i \Phi}, e^{i \phi})$, 
an annulus of area $p$, and another disc, obtaining
\begin{equation}
Z_\YM^\open(N, g_s, \theta, \phi) = \IP{\Psi_\phi | A_p | \Psi_0}.
\end{equation}
Using the factorization results \eqref{annulus-factorized}, \eqref{disc-factorized} and \eqref{dual-wilson-line-freeze-factorized} this becomes
\begin{multline} \label{ym-sphere-dual-freeze-factorized}
Z_\YM^\open(N, g_s, \theta, \phi) = Z^0_\YM(N, g_s, \theta, \phi) M(q)^2 \eta(q)^{2N} \times \\
\sum_{l, m \in \Z, R'_1, R'_2} (-)^{\abs{R_1}+\abs{R_2}} q^{\half p l^2} e^{iNlp\theta} \det(e^{im\phi}) Z^{R'_1, R'_2,l,m}_+ Z^{R'_1, R'_2,l,m}_- + \OO(e^{-N}),
\end{multline} 
with
\begin{multline} \label{chiral-block-sphere-dual-freeze}
Z^{R'_1, R'_2,l,m}_+(N, g_s, \theta, \phi) = q^{\half N(\abs{R'_1}+\abs{R'_2})} \sum_{R_+, T_+} q^{\half p \kappa_{R_+} + \half \kappa_{T_+} + (\half N(p-2)+lp-m) \abs{R_+} + (- \half N-l) \abs{T_+}} \times \\
C_{T_+ R'_1 R_+} C_{R'^t_2 R_+ 0} \schur_{T^t_+} (e^{-i\phi}) e^{i \theta p \abs{R_+}},
\end{multline}
and similarly
\begin{multline} \label{antichiral-block-sphere-dual-freeze}
Z^{R'_1, R'_2,l,m}_-(N, g_s, \theta, \phi) = q^{\half N(\abs{R'_1}+\abs{R'_2})} \sum_{R_-, T_-} q^{\half p \kappa_{R_-} + \half \kappa_{T_-} + (\half N(p-2)-lp+m) \abs{R_-} + (- \half N + l)\abs{T_-}} \times \\
C_{T_- R'^t_1 R_-} C_{R'_2 R_- 0} \schur_{T^t_-} (e^{i\phi}) e^{-i \theta p \abs{R_-}}.
\end{multline}

\begin{figure}[t]
\centering
\psset{unit=0.35}
\input{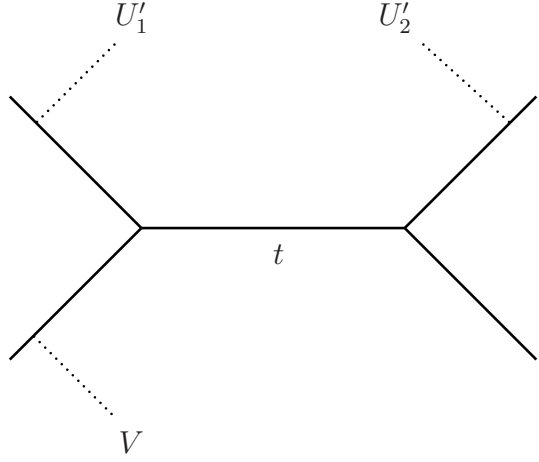}
\caption{The vertex representation of $X$, with a stack of $M$ Lagrangian branes with
complexified holonomy $V = e^v$, a stack of infinitely many ghost branes with complexified holonomy $U'_1 = e^{u'_1}$,
and a stack of infinitely many ghost antibranes with complexified holonomy $U'_2 = e^{u'_2}$.}
\label{fig-toric-external}
\end{figure}

As in case 2, we can now interpret these chiral blocks in terms of the topological string on $X$ with $M$ Lagrangian branes, now inserted on the external leg as indicated in Figure \ref{fig-toric-external}.  Again from \cite{Aganagic:2003db}, the
topological partition function in this geometry (with a particular choice of framing on the $M$ external branes) is
\begin{multline} \label{top-sphere-external-line}
\psi_\Top^{\ghost}(g_\Top,t,v,u') = \psi_\Top^0(g_\Top,t,v) 
\sum_{R,T} C_{T R'_1 R} C_{R'_2 R 0} \schur_{R'_1}(e^{u'_1}) \schur_{R'_2}(e^{u'_2}) \times \\
 (-)^{p \abs{R}} q^{\half p \kappa_{R} + \half \kappa_T} e^{-t \abs{R}} \schur_{T^t}(e^{-v}),
\end{multline}
and similarly one can compute $\psi_\Top^\antighost$ with the ghost branes exchanged for antibranes.
Now define
\begin{align}
t &= \half N g_s(p-2) - i \theta p, \\
u'_1 &= \half N g_s + i \phi'_1, \\
u'_2 &= \half N g_s + i \phi'_2, \\
v &=  - \half N g_s + i \phi, \\
g_\Top &= g_s.
\end{align}
With this substitution and a suitable normalization one can relate \eqref{top-sphere-external-line} to the chiral blocks
appearing in the factorization \eqref{ym-sphere-dual-freeze-factorized}, similarly to what was done above in case 2,
obtaining
\begin{multline} \label{2dym-factorization-open-fiber}
Z_{\YM}(N, g_s, \theta, \phi) = \sum_{l,m \in \Z} \int d_H \phi'_1 d_H \phi'_2 \times \\ 
\left(\psi_\Top^{\ghost} \left(g_s, t + l p g_s, v - l g_s, u'_1 - m g_s, u'_2 \right) \right) \overline{\left(\psi_\Top^{\antighost} \left(g_s, t - l p g_s, v + l g_s, u'_1 + m g_s, u'_2  \right) \right)}.
\end{multline}

So finally we have found that $Z_\YM^\open = \abs{\psi_\Top^\open}^2 + \OO(e^{-N})$, both for branes in the fiber
and in the base.

\subsection*{Summing open D2-branes on $\OO(-1) \oplus \OO(-1) \to \C\PP^1$}

Next we want to verify that $Z_\YM^\open$ can indeed be interpreted as counting open D2-branes
with the chemical potential $\varphi_e^\open = 2 \pi \phi / g_s$.
We consider case $2$, where we have a Wilson line
freezing operator $\delta_M\left(e^{i \oint_{\gamma} A}, e^{i \phi}\right)$ 
cutting the sphere into two pieces, with discrete areas $p_1$, $p_2$ such that $p_1 + p_2 = p$,
and further specialize to the case $p_1 = p_2 = 1$.

As we did in the previous section, we compute the partition function $Z_\YM^\open$ of this Yang-Mills theory
using the gluing procedure described in Appendix \ref{app-qym}:  namely, we construct the sphere by
gluing two discs together with the operator $\delta_M(e^{i \oint A}, e^{i \phi})$ in the middle.
However, unlike above where we used the splitting $\HH = \HH_+ \otimes \HH_-$ to see the large $N$ factorization
of $Z_\YM^\open$, in this section we will write the explicit formula for $Z_\YM^\open$ at finite $N$.

The wave function of the $q$-deformed 2-dimensional Yang-Mills theory on the disc is a function of the eigenvalues
$\xi$ of the Wilson line around the boundary, evaluated in Appendix \ref{app-disc-wave-function}:
\begin{equation}
\Psi(\xi) = e^{- N g_s / 24} \Theta_N \left(\frac{1}{2\pi}(\xi + \theta), \frac{i g_s}{2\pi} \right),
\end{equation}
where $\Theta_N$ denotes the theta function of $\Z^N$,
\begin{equation} \label{theta-function-redux}
\Theta_N (z,\tau) = \sum_{\gamma \in \Z^N} e^{\pi i \tau \norm{\gamma}^2} e^{2 \pi i \IP{\gamma,z}}, \qquad \for z \in \R^N,\,\im \tau > 0.
\end{equation}
In our case we want to glue two such disc wave functions $\Psi_1(\xi)$, $\Psi_2(\xi)$ 
to one another with $\delta_M(e^{i \oint A}, e^{i \phi})$ sandwiched in the middle.  
The result of this gluing is given in \eqref{wilson-line-freeze}, but we need a little notation first:
we divide the lattice $\Z^N$ into $\Z^{N-M} \oplus \Z^M$, and correspondingly divide $\xi$ into $\zeta$ and $\phi$, with $N-M$ and $M$ components respectively.  Then the result
of the gluing is
\begin{equation} \label{integral-pairing}
Z_\YM^\open(N, g_s, \theta, \phi) = \int_{[0,2\pi]^{N-M}} \frac{d\zeta}{2\pi} \abs{D(\zeta)}^2 \Psi_1(-\zeta,-\phi) \Psi_2(\zeta,\phi).
\end{equation} 
Because of the simple form of the wave function, the $\zeta$ dependence and $\phi$ dependence decouple, namely
\begin{equation}
\Psi(\zeta,\phi) = e^{- N g_s / 24} \Theta_{N-M} \left(\frac{1}{2\pi}(\zeta + \theta), \frac{i g_s}{2\pi} \right) \Theta_M \left(\frac{1}{2\pi}(\phi + \theta), \frac{i g_s}{2\pi} \right).
\end{equation}
So write
\begin{equation}
f_{N-M}(\theta, g_s) = \int \frac{d\zeta}{2\pi} \abs{D(\zeta)}^2 \Theta_{N-M} \left(\frac{1}{2\pi}(\zeta + \theta), \frac{i g_s}{2\pi} \right) \Theta_{N-M} \left(\frac{1}{2\pi}(-\zeta + \theta), \frac{i g_s}{2\pi} \right).
\end{equation} 
Then \eqref{integral-pairing} becomes
\begin{equation}
Z_\YM^\open(N, g_s, \theta, \phi) = e^{- N g_s / 12} f_{N-M}(\theta, g_s) \Theta_M \left(\frac{1}{2\pi}(\phi + \theta), \frac{i g_s}{2\pi} \right) \Theta_M \left(\frac{1}{2\pi}(- \phi + \theta), \frac{i g_s}{2\pi} \right).
\end{equation} 
Now we can use the Poisson resummation property of the theta function,
\begin{equation}
\Theta_M(z,\tau) = \left(\frac{i}{\tau}\right)^{M/2} e^{-\pi i \norm{z}^2 / \tau} \Theta_M(z / \tau, - 1 / \tau),
\end{equation} 
to obtain
\begin{multline}
Z_\YM^\open(N, g_s, \theta, \phi) = e^{- N g_s / 12} f_{N-M}(\theta, g_s) \left( \frac{2\pi}{g_s} \right)^M e^{-\frac{1}{2g_s} (\norm{\phi + \theta}^2 + \norm{\phi - \theta}^2)} \times \\
\Theta_M \left(-\frac{i}{g_s}(\phi + \theta), \frac{2 \pi i}{g_s} \right) \Theta_M \left(-\frac{i}{g_s}(- \phi + \theta), \frac{2 \pi i}{g_s} \right).
\end{multline} 
Expanding out these theta functions then gives 
$Z_\YM^\open(N,g_s,\theta,\phi)$ as an expansion in $e^{- 2 \pi \phi / g_s}$, up to a prefactor
$e^{-\frac{1}{g_s} \norm{\phi}^2}$.
So up to this prefactor, we have verified that $Z_\YM^\open$ can indeed be interpreted as counting open D2-branes
with the chemical potential $\varphi_2^\open = 2 \pi \phi / g_s$.

For completeness, let us briefly consider the leftover factor $f_{N-M}(\theta,g_s)$.  Writing out using \eqref{vandermonde}
\begin{equation}
\abs{D(\zeta)}^2 = \sum_{\sigma, \sigma' \in S_{N-M}} (-)^{\sigma\sigma'} e^{i \IP{\zeta, \sigma(\rho)-\sigma'(\rho)}}
\end{equation} 
(where $\rho = \rho_{N-M}$) and evaluating the integral using the definitions of the theta functions gives
\begin{equation}
f_{N-M}(\theta,g_s) =
\sum_{\sigma,\sigma' \in S_{N-M}} (-)^{\sigma \sigma'} e^{- \half g_s \norm{\sigma(\rho) - \sigma'(\rho)}^2} \Theta_{N-M} \left( \frac{1}{2\pi} (-2 \theta + i g_s (\sigma(\rho) - \sigma'(\rho))), \frac{i g_s}{\pi} \right).
\end{equation} 
So this can also be resummed to give an expansion in $e^{-4 \pi^2 / g_s}$ and $e^{-2 \pi \theta / g_s}$, as one expects from the
closed string sector of the conjecture.

\section*{Acknowledgements}
We would like to thank Jacques Distler, Noam Elkies, Sergei Gukov, Marcos Mari\~{n}o, Shiraz Minwalla, Lubo\v{s} Motl, Hirosi Ooguri and Natalia Saulina for useful discussions.  The research of M.A. was supported in part by a DOE OJI Award and an Alfred P. Sloan Foundation fellowship.  The research of A.N. and C.V. was supported in part by NSF grants PHY-0244821 and DMS-0244464.

\appendix

\section{Group theory} \label{app-group-theory}

In this appendix we summarize our group theory conventions and a few useful formulas.

We use script letters $\cR, \cP, \cQ, \dots$ to denote representations of unitary groups
such as $U(N)$, and capital letters $R, P, Q, \dots$
to denote Young diagrams.  Often Young diagrams will appear as the chiral and anti-chiral parts $R_\pm$
of a representation $\cR = R_+ \overline{R_-}[l]$, as described in Appendix \ref{app-factorization}.

The weight lattice of $U(N)$ is $\Z^N$, with its standard inner product $\IP{,}$.  
A highest weight representation $\cR$ is characterized by a
weight $(r_1, \dots, r_N) \in \Z^N$, in a particular Weyl chamber; we make the standard choice of Weyl
chamber, given by the constraint $r_1 \ge \cdots \ge r_N$.  With this choice, the entries $r_i$ correspond
to the lengths of the rows of the extended Young diagram for the representation $\cR$.
The Weyl group $W$ of $U(N)$ is the symmetric group, $W \simeq S_N$,
which permutes the entries of $\Z^N$ in the obvious way.  

We will use the symbol $\cR$ both
for the representation and for its highest weight.
It is also convenient to
introduce the symbol $\hat{\cR}$ for $\cR + \rho$, where $\rho$ is half the sum of the positive roots of $U(N)$,
concretely
\begin{equation} \label{rho}
\rho = \half(N-1, N-3, \dots, 3-N, 1-N).
\end{equation}
We also write $\bo$ for the ``unit'' vector,
\begin{equation} \label{bo}
\bo = (1, 1, \dots, 1, 1).
\end{equation}
With this notation we can write the Weyl character formula,\footnote{When $N$ is odd, 
$D(\xi)^{-1}$ and $\sum_{\sigma \in S_N} (-)^\sigma e^{i\IP{\hat{\cR}, \sigma(\xi)}}$ are not
quite well defined as functions of the eigenvalues $e^{i\xi}$ --- they change sign under $\xi_i \to \xi_i + 2\pi$.  Nevertheless their product is still well defined.}
\begin{equation} \label{weyl-character-formula}
\Tr_\cR(e^{i \xi}) = D(\xi)^{-1} \sum_{\sigma \in S_N} (-)^\sigma e^{i\IP{\hat{\cR}, \sigma(\xi)}},
\end{equation}
where the denominator $D(\xi)$ is
\begin{equation} \label{vandermonde}
D(\xi) = \sum_{\sigma \in S_N} (-)^\sigma e^{i\IP{\xi, \sigma(\rho)}} = \prod_{i<j} (e^{i (\xi_i - \xi_j) / 2} - e^{- i (\xi_i - \xi_j) / 2}).
\end{equation}

In computing the $q$-deformed Yang-Mills amplitudes we will need to use the Hopf link invariant 
$S_{\cP \cQ}$ of the level $k$ Chern-Simons theory with gauge group $U(N)$.  Define $g_s = \frac{2\pi}{N+k}$.
There is a formula expressing $S_{\cP \cQ}$ as a sum over the Weyl group $W \simeq S_N$:
\begin{equation} \label{s-matrix-formula}
S_{\cP \cQ} = e^{- g_s (\norm{\rho}^2 + N/24)} \sum_{\sigma \in S_N} (-)^\sigma e^{g_s \IP{\hat{\cP}, \sigma(\hat{\cQ})}}.
\end{equation}
(The standard formulas for $S_{\cP \cQ}$ include a different normalization, 
but in the context where we will use $S_{\cP \cQ}$ we will absorb this in 
other normalization factors.)

For any $N_1, N_2$ with $N_1 + N_2 = N$, let $\cQ$ label a representation of $U(N_1)$ and $\cA$ a representation of $U(N_2)$, while $\cR$ is a representation of $U(N)$; 
then we define the branching coefficients $\cB^{\cR}_{\cQ \cA}$ by 
the rule that $\cR$ decomposes under $U(N_1) \times U(N_2)$ as
\begin{equation} \label{branching}
\cR \to \bigoplus_{\cQ,\cA} \cB^{\cR}_{\cQ \cA} [\cQ, \cA].
\end{equation}

We fix the normalization of the Casimir operators of $U(N)$ 
as follows:  in a representation $\cR$ with highest weight $(r_1, \dots, r_N)$,
\begin{align}
C_1(\cR) &= \IP{\hat\cR, \bo} = \sum_i r_i, \label{c1} \\
C_2(\cR) &= \norm{\hat\cR}^2 - \norm{\rho}^2 = \sum_i r_i (r_i + N + 1 - 2i). \label{c2}
\end{align}

We write $N^R_{R_1 R_2}$ for the usual Littlewood-Richardson numbers, and also use a slight
generalization which we write $N^R_{R_1 \cdots R_k}$.  These numbers can be defined
in various equivalent ways --- for example, if we think of the Young diagrams $R_i$ and $R$ as
representations of $GL(\infty)$, they are the tensor product coefficients, i.e.
\begin{equation}
R_1 \otimes \cdots \otimes R_k = \bigoplus_{R} N^R_{R_1 \cdots R_k} R.
\end{equation}
In particular, $N^R_{R_1 \cdots R_k} = 0$  unless $\sum_{i=1}^k \abs{R_i} = \abs{R}$,
where $\abs{R}$ denotes the total number of boxes in the diagram $R$.

We write $\schur_R(x)$ for the ``Schur function'' associated with the Young diagram $R$:  this
is a symmetric polynomial in infinitely many variables, $x = (x_1, x_2, \dots)$.  It can be
defined in various equivalent ways; one convenient way to think of it is as the character
of the $\mathrm{Mat}(\infty,\C)$ representation associated to $R$, evaluated on the diagonal matrix with
entries $(x_1, x_2, \dots)$.  There is a bilinear inner product $\IP{,}$ on the ring of symmetric polynomials 
for which the Schur functions form an orthonormal basis, $\IP{\schur_R, \schur_S} = \delta_{RS}$; 
in terms of this inner product $N^R_{R_1 \cdots R_k} = \IP{\prod_{i=1}^k \schur_{R_i}, \schur_R}$. 
Viewing the $x_i$ as eigenvalues, the inner product can be written as a formal integral of class 
functions over $U(\infty)$ (interpreted as an inverse limit of finite-dimensional groups 
with their normalized Haar measures),
\begin{equation}
\IP{f,g} = \int d_H \xi\,f(e^{-i\xi}) g(e^{i\xi}).
\end{equation}
We also use the ``skew Schur functions'' $\schur_{R/A}(x)$, defined by
\begin{equation} \label{skew-schur-function}
\schur_{R/A}(x) = \sum_Q N^R_{Q A} \schur_Q(x).
\end{equation}
See \cite{MR1354144} for much more on Schur functions and skew Schur functions.

We also introduce an analog of the skew Schur function, a ``skew trace''
involving the branching $U(N) \to U(N_1) \times U(N_2)$ where $N_1 + N_2 = N$:  this is a rule by
which a representation of $U(N)$ and a representation of $U(N_2)$ induce
a class function on $U(N_1)$, which we define by
\begin{equation} \label{skew-trace}
\Tr_{\cR / \cA}(U) = \sum_\cQ \cB^\cR_{\cQ \cA} \Tr_\cQ (U).
\end{equation} 
Here $\cR, \cQ, \cA$ denote representations of $U(N)$, $U(N_1)$, $U(N_2)$ respectively;
$\cB$ denotes the branching coefficients defined in \eqref{branching}; and $U \in U(N_1)$.

We will frequently encounter sums $\sum_{R'}$ over the set of all Young diagrams.  A particularly
useful identity for reducing such sums is
\begin{equation} \label{ghost-cancellation}
\sum_{R',S'} (-)^{\abs{R'}} N_{R' S' A_{1} \cdots A_{a}}^{A} N_{R'^t S' B_1 \cdots B_b}^{B} = N_{A_1 \cdots A_a} ^A N_{B_1 \cdots B_b}^B.
\end{equation}
One can prove \eqref{ghost-cancellation} using the ``Cauchy identities'' for Schur functions, given e.g. in \cite{MR1354144},
\begin{align}
\sum_{S'} \schur_{S'}(x) \schur_{S'}(y) &= \prod_{i=1}^\infty \prod_{j=1}^\infty (1 - x_i y_j)^{-1},\\
\sum_{R'} (-)^{\abs{R'}} \schur_{R'}(x) \schur_{R'^t}(y) &= \prod_{i=1}^\infty \prod_{j=1}^\infty (1 - x_i y_j).
\end{align}

\section{The $q$-deformed 2-d Yang-Mills theory} \label{app-qym}

In this section we review some facts about 2-dimensional Yang-Mills theory and its $q$-deformed cousin.
We begin with the 2-dimensional Euclidean Yang-Mills action\footnote{Note that our convention for $\theta_{\YM}$ is not the usual one; $\theta_\YM^{\mathrm{usual}} = \frac{i \pi}{g^2_\YM} \theta_\YM$.} for gauge group $G = U(N)$,
\begin{equation} \label{ym-action}
S_{\YM} = \frac{1}{2 g^2_{\YM}} \left( \int_\Sigma d^2x\ \Tr F \wedge *F + \theta_{\YM} \int_\Sigma \Tr F \right).
\end{equation}
It is often convenient to rewrite \eqref{ym-action} by introducing an additional adjoint-valued scalar field $\Phi$, which enters the
action quadratically:  namely, \eqref{ym-action} is equivalent to
\begin{equation}
S_{\YM} = \frac{1}{2 g^2_\YM} \left( 2 \int_\Sigma \Tr \Phi F - \int_\Sigma \mu\ \Tr \Phi^2 + \theta_{\YM} \int_\Sigma \mu\ \Tr \Phi \right),
\end{equation}
where $\mu$ is the area element on $\Sigma$.
Once we have introduced this $\Phi$ we can define the $q$-deformed theory:  we use the same
action $S_{\YM}$, but we consider the fundamental variables to be the gauge connection and $e^{i \Phi}$,
rather than the gauge connection and $\Phi$.  More precisely, since there is an ambiguity in recovering
$\Phi$ from $e^{i \Phi}$, $S_{\YM}$ is not well defined as a function of $\Phi$; to get a well
defined expression inside the path integral one has to sum $e^{-S_{\YM}}$ over all ``images'' $\Phi$.  
Equivalently, we integrate over all $\Phi$, not just a fundamental domain, 
but we use the measure appropriate for an integral over $e^{i \Phi}$.
This construction gives the $q$-deformed theory with $q = e^{- g^2_\YM}$, which is the one
that naturally occurs in this paper; to get a different value of $q$
one would change the periodicity of $\Phi$.

The partition function can be computed in various ways; here we will focus on the computation by cutting and pasting.  In the case of the undeformed Yang-Mills theory, this procedure was reviewed in \cite{Cordes:1995fc}; our treatment will be briefer, and is 
intended mostly to recall the new features that appear in the $q$-deformed case, as
described in \cite{Aganagic:2004js}.

To get the cutting-and-pasting procedure started one first needs to 
know the Hilbert space $\HH$ of the theory on $S^1$; as for the usual 2-d Yang-Mills theory, it is simply the space of class functions
$\Psi(g)$, with $g \in G$ interpreted as the holonomy of the connection around $S^1$.  The path integral over a surface with
boundary $S^1$ thus gives a state $\Psi(g)$.  Two such surfaces can be glued using the rule\footnote{We will always use the notation
$\IP{|}$ to stand for the (linear) gluing rule rather than the (sesquilinear) inner product on the Hilbert space.  The two are the same
when acting on real linear combinations of the characters $\Tr_\cR(g)$ but differ for complex linear combinations.}
\begin{equation}
\IP{\Psi_1 \vert \Psi_2} = \int_G d_H g\,\Psi_1(g^{-1}) \Psi_2(g),
\end{equation}
with $d_H g$ the Haar measure.  When $G = U(N)$ we can write these wave functions 
more concretely as functions of the eigenvalues $e^{i \xi_i}$,
totally symmetric under the permutation group $S_N$, and the gluing rule becomes
\begin{equation} \label{gluing-rule}
\IP{\Psi_1 \vert \Psi_2} = \int_{[0,2\pi]^N} \frac{d\xi}{2\pi} \abs{D(\xi)}^2 \, \Psi_1(-\xi) \Psi_2(\xi),
\end{equation}
where as in \eqref{vandermonde}
\begin{equation} \label{vandermonde-redux}
D(\xi) = \prod_{i<j} (e^{i (\xi_i - \xi_j) / 2} - e^{- i (\xi_i - \xi_j) / 2}).
\end{equation}
A convenient basis of $\HH$ (which in particular diagonalizes the Hamiltonian) is given by the characters $\Tr_\cR(g)$ as $\cR$ runs over all representations of $G$.  In that basis the gluing rule becomes
\begin{equation}
\IP{\cR_1 \vert \cR_2} = \delta_{\cR_1 \cR_2}.
\end{equation} 

Next we need the partition function on a few elementary surfaces, from which any $\Sigma$ of interest to us can be pasted together:

\medskip

\noindent {\bf The annulus.}  The annulus of area $a$ has two boundaries, so it gives an operator $A_a: \HH \to \HH$.  The gluing
rule for an annulus can be obtained directly from the action by working out the Hamiltonian; it is
\cite{Aganagic:2004js}
\begin{equation} \label{annulus}
\IP{\cR_1 \vert A_a \vert \cR_2} = \delta_{\cR_1 \cR_2} e^{- a \left(\half g^2_{\YM} C_2(\cR) - i \theta_\YM C_1(\cR) \right)}.
\end{equation}

\medskip 

\noindent {\bf The Wilson line freezing operator.} 
As discussed in Section \ref{sec-open}, we will be particularly interested in computing amplitudes 
involving a particular operator, written $\delta_M(e^{i \oint A}, e^{i \phi})$,
which has the effect of freezing $M$ of the eigenvalues along a Wilson loop to
the values $e^{i\phi_1}, \dots, e^{i\phi_M}$.
A natural guess for the gluing rule with $\delta_M(e^{i \oint A}, e^{i \phi})$ inserted can be obtained by splitting up
the $N$ eigenvalues $\xi_i$ into $(\underbrace{\zeta}_{N-M}, \underbrace{\phi}_M)$:  namely, one 
freezes the $\phi$ eigenvalues in the gluing rule \eqref{gluing-rule} and integrates only over the
$\zeta$ eigenvalues, obtaining
\begin{equation} \label{wilson-line-freeze}
\IP{\Psi_1 \vert \delta_M(e^{i \oint A}, e^{i \phi}) \vert \Psi_2} = \int_{[0,2\pi]^{N-M}} \frac{d\zeta}{2\pi} \abs{D(\zeta)}^2 \,\, 
\Psi_1(-\zeta, -\phi) \Psi_2(\zeta, \phi).
\end{equation}
This integral has an interpretation in the representation basis.  Namely, suppose $\Psi_j(\xi) = 
\Tr_{\cR_j}(e^{i \xi})$.  Then decomposing $\cR_j$ under $U(N-M) \times U(M)$ gives
\begin{equation}
\Psi_j(\xi) = \sum_{\cA_j,\cQ_j} \cB^{\cR_j}_{\cA_j \cQ_j} \Tr_{\cA_j}(e^{i \zeta}) \Tr_{\cQ_j}(e^{i\phi}).
\end{equation} 
The integral over $\zeta$ then picks out the terms with $\cA_1 = \cA_2$, giving
\begin{equation}
\IP{\cR_1 \vert \delta_M(e^{i \oint A}, e^{i \phi}) \vert \cR_2} = \sum_{\cQ_1, \cQ_2, \cA} \cB^{\cR_1}_{\cA \cQ_1} \cB^{\cR_2}_{\cA \cQ_2} \Tr_{\cQ_1}(e^{-i \phi}) \Tr_{\cQ_2}(e^{i \phi}).
\end{equation} 
If we define the skew trace $\Tr_{\cR / \cS}$ as in \eqref{skew-trace},
we can rewrite this as
\begin{equation} \label{wilson-line-freeze-rep-basis}
\IP{\cR_1 \vert \delta_M(e^{i \oint A}, e^{i \phi}) \vert \cR_2} = \sum_{\cA} \Tr_{\cR_1 / \cA} (e^{-i\phi}) \Tr_{\cR_2 / \cA} (e^{i\phi}).
\end{equation}

\medskip

\noindent {\bf The disc.}
The disc of zero area gives a simple state $\Psi_0 \in \HH$ on its boundary,
\begin{equation} \label{disc}
\Psi_0 = \sum_{\cR} S_{\cR 0} \ket{\cR},
\end{equation}
as was computed in \cite{Aganagic:2004js}.  (This should be compared to the 
analogous expression in the non-$q$-deformed Yang-Mills theory --- there one would 
have replaced $S_{\cR 0}$ by $\dim \cR$ up to overall normalization.  
Indeed, $S_{\cR 0} / S_{00}$ is the quantum dimension
$\dim_q \cR$.)

\medskip

\noindent {\bf The dual Wilson line freezing operator.}
Also as discussed in Section \ref{sec-open}, we need the operator $\delta_M(e^{i \Phi}, e^{i \phi})$ which freezes
$M$ of the eigenvalues of the dual Wilson line $\Phi$ at a point of $\Sigma$.  The disc of zero area with this operator 
inserted gives a state $\Psi_\phi \in \HH$ on its boundary, for which the natural formula is
\begin{equation} \label{dual-wilson-line-freeze}
\Psi_\phi = \sum_{\cR,\cS} S_{\cR \cS} \Tr_{\cS / 0} (e^{i \phi}) \ket{\cR}.
\end{equation}
This is a straightforward generalization of the result of \cite{Aganagic:2004js} in the case
$M=N$, along the lines of what we did above for the Wilson line freezing operator.  (In 
the special case $M=N$, the result of \cite{Aganagic:2004js} just 
replaces $\Tr_{\cS / 0}$ by $\Tr_\cS$ in the above.)

\medskip

\noindent {\bf The trinion (pair of pants).}  The trinion has three boundaries, so it gives an element
in $\HH \otimes \HH \otimes \HH$, namely
\begin{equation} \label{trinion}
T = \sum_\cR \frac{\lvert \cR \rangle \otimes \lvert \cR \rangle \otimes \lvert \cR \rangle}{S_{0\cR}}.
\end{equation}
It was computed in \cite{Aganagic:2004js}; we include it here just for 
completeness since it would be relevant for Riemann surfaces of genus $g>1$.

\medskip

In addition to these local ingredients we will include an overall normalization factor $Z_\YM^0$ in the partition function of this $q$-deformed theory; we do not give a general rule for this normalization
here, but in the example we consider in the text, it can be found in \eqref{ym-normalization-sphere}.

A $q$-deformation of 2-dimensional Yang-Mills theory has also been considered in \cite{Buffenoir:1994fh}
where it was formulated using a lattice regularization.  That formulation is likely to be equivalent to the one discussed here and in \cite{Aganagic:2004js}; at least the partition function appears to be the same on an arbitrary surface.

\section{The disc wave function} \label{app-disc-wave-function}

Consider the $q$-deformed 2-d Yang-Mills theory on a disc of area $p$, with parameters fixed by
\begin{equation} \label{2dym-params-redux2}
\theta_{\YM} = \theta, \quad g^2_{\YM} = g_s, \quad q = e^{-g_s}.
\end{equation}
The path integral on this disc gives a state $\Psi(\xi)$ on the boundary, for which one
can give a formula using the rules of Appendix \ref{app-qym}; namely, it is a sum over
irreducible representations $\cR$ of $U(N)$,
\begin{equation} \label{disc-wavefunction}
\Psi(\xi) = \sum_\cR S_{\cR 0} e^{- \half p g_s C_2(\cR)} e^{i \theta p C_1(\cR)} \Tr_\cR e^{i\xi}. 
\end{equation}
Our purpose in this section is to express this $\Psi(\xi)$ in terms of theta functions.  
As reviewed in Appendix \ref{app-group-theory}, the irreducible representations of $\cR$ can be labeled by
their highest weights $\cR = (r_1, \dots, r_N) \in \Z^N$, subject to the constraint $r_1 \ge r_2 \ge \cdots \ge r_N$.
We also write $\hat\cR = \cR + \rho$.  Now we use the Weyl character formula \eqref{weyl-character-formula},
the modular S matrix formula \eqref{s-matrix-formula}, and the Casimirs \eqref{c1}, \eqref{c2};
altogether \eqref{disc-wavefunction} becomes
\begin{multline}
\Psi(\xi) = \sum_\cR \left( e^{-g_s(\norm{\rho}^2 + N/24)} \sum_{\sigma \in S_N} (-)^\sigma e^{g_s \IP{\sigma(\hat{\cR}), \rho}} \right) e^{- \half p g_s (\norm{\hat{\cR}}^2 - \norm{\rho}^2) + i \theta p \IP{\hat{\cR},\bo}} \times \\
\left( D(\xi)^{-1} \sum_{\sigma' \in S_N} (-)^{\sigma'} e^{i \IP{\sigma'(\hat{\cR}), \xi}} \right).
\end{multline}
Writing $\tilde{\sigma} = \sigma \sigma'^{-1}$, and using the Weyl invariance of $\IP{,}$ and $\bo$, 
we can rewrite this as
\begin{equation}
\Psi(\xi) = e^{-g_s(\norm{\rho}^2 + N/24)} D(\xi)^{-1} e^{\half p g_s \norm{\rho}^2} \sum_\cR \sum_{\sigma,\tilde\sigma \in S_N} (-)^{\tilde\sigma} e^{-\half p g_s \norm{\hat{\cR}}^2 + i \IP{\sigma(\hat{\cR}), \tilde{\sigma}(\xi) + \theta p \bo - i g_s \rho}}.
\end{equation}
Now we want to express this as a theta function.  If $\cR$ runs over all weight vectors in a given Weyl chamber, then it is easy to see that $\hat\cR$ runs over all weight vectors in the \ti{interior} of that chamber.\footnote{If $N$ is even, the weight lattice has to be shifted by $\half \bo$.  This subtlety modifies some of our intermediate expressions but cancels out in the final result
\eqref{disc-wavefunction-simplified}.}  Since the Weyl group acts transitively to permute the Weyl chambers, the sum over $\sigma$ and $\cR$ can be combined into a single sum over $\gamma = \sigma(\hat{\cR})$, where $\gamma$ runs over the weight lattice $\Z^N$, or more precisely over those vectors in $\Z^N$ which are not in the boundary of any Weyl chamber.  In terms of $\gamma$ the sum becomes
\begin{equation}
\Psi(\xi) = e^{-g_s(\norm{\rho}^2 + N/24)} D(\xi)^{-1} e^{\half p g_s \norm{\rho}^2} \sum_\gamma e^{-\half p g_s \norm{\gamma}^2 + i\IP{\gamma, \theta p \bo - i g_s \rho} } \sum_{\tilde\sigma \in S_N} (-)^{\tilde\sigma} e^{i \IP{\gamma, \tilde{\sigma} (\xi)}}.
\end{equation}
But now note that the sum over $\tilde{\sigma}$ vanishes if $\gamma$ is
fixed by some Weyl reflection $\tilde{\sigma}$, i.e. if it lies on the boundary of a Weyl chamber.
Therefore we can extend the sum over $\gamma$ to run over the whole weight lattice $\Z^N$.  The sum can be written (now dropping the $\tilde{}$ on $\sigma$ for notational simplicity)
\begin{equation} \label{disc-wavefunction-theta}
\Psi(\xi) = e^{-g_s(\norm{\rho}^2 + N/24)} D(\xi)^{-1} e^{\half p g_s \norm{\rho}^2} \sum_{\sigma \in S_N} (-)^\sigma \Theta_N \left(\frac{1}{2\pi}(\sigma(\xi) + \theta p \bo - i g_s \rho), \frac{i p g_s}{2\pi} \right).
\end{equation}
Here we have introduced the theta function of $\Z^N$,
\begin{equation} \label{theta-function}
\Theta_N (z,\tau) = \sum_{\gamma \in \Z^N} e^{\pi i \tau \norm{\gamma}^2} e^{2 \pi i \IP{\gamma,z}}, \qquad \for z \in \R^N,\,\im \tau > 0,
\end{equation}
which obeys the functional equation
\begin{equation} \label{theta-functional-equation}
\Theta_N (z+\tau \lambda, \tau) = e^{- \pi i \tau \norm{\lambda}^2} e^{-2 \pi i \IP{\lambda,z}} \Theta_N(z,\tau) \qquad \for \lambda \in \Z^N.
\end{equation}
One can simplify \eqref{disc-wavefunction-theta} in the case $p=1$; namely,
in this case, one can apply \eqref{theta-functional-equation} with $\lambda = -\rho$.
After some straightforward algebra using \eqref{vandermonde} the sum over $\sigma$ cancels the denominator $D(\xi)$, leaving
\begin{equation} \label{disc-wavefunction-simplified}
\Psi(\xi) = e^{-g_s N/24} \Theta_N \left(\frac{1}{2\pi}(\xi + \theta \bo), \frac{i g_s}{2\pi} \right).
\end{equation}

\section{Factorization} \label{app-factorization}

In this appendix we give some mathematical results which are used in the text to establish
the factorization of the 2-dimensional Yang-Mills amplitude with operator insertions into chiral and anti-chiral parts.

\subsection*{Coupled representations}

An essential technical tool in studying the factorization of 2-d Yang-Mills into chiral and anti-chiral sectors, introduced
in \cite{Gross:1993hu}, is the notion of a \ti{coupled representation} of $U(N)$.  Here we review the notion of coupled
representation.

\begin{figure}[t]
\centering
\psset{unit=0.60}
\input{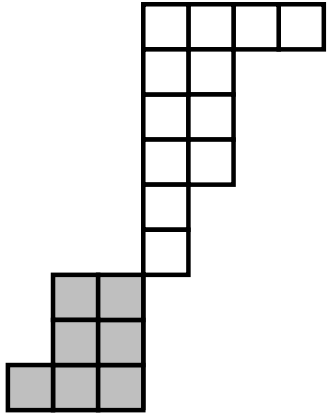}
\caption{An extended Young diagram representing a representation of $U(N)$ (for $N=9$) constructed by symmetrization and antisymmetrization over both fundamental representations (white boxes) and antifundamental representations (grey boxes).}
\label{fig-young-antiboxes}
\end{figure}

Recall that the irreducible representations of $SU(N)$ correspond to Young diagrams with no more than $N$ rows.
Such a diagram can be specified by giving the lengths of the rows, $(\lambda_1, \dots, \lambda_N)$, with $\lambda_1 \ge
\lambda_2 \ge \cdots \ge \lambda_N$, and all $\lambda_i \ge 0$.  Denote the fundamental representation by $V$.  Then
the representation of $SU(N)$ corresponding to $\lambda$ is obtained as a particular subspace of $V^{\otimes \abs{\lambda}}$, roughly by symmetrizing over the rows and antisymmetrizing over the columns.
In the case of $SU(N)$ one can obtain all representations in this way, but for $U(N)$
one also needs to include copies of the antifundamental representation $\overline{V}$.  This corresponds to considering ``extended Young diagrams''
which can include ``antiboxes'' as well as boxes, i.e. removing the constraint that all $\lambda_i \ge 0$, as shown in Figure \ref{fig-young-antiboxes}.

\begin{figure}[t]
\centering
\psset{unit=0.55}
\input{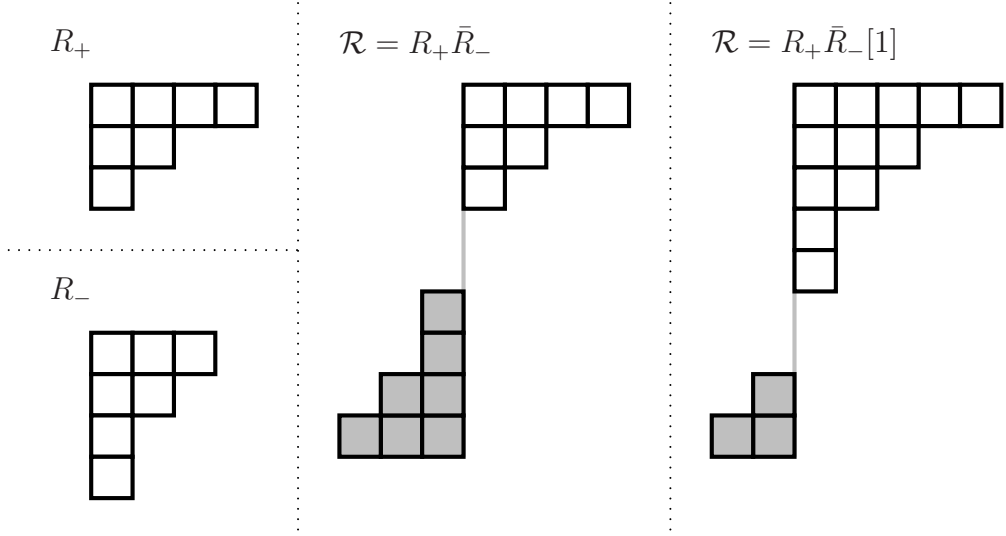}
\caption{A coupled representation of $U(N)$ (for $N=9$).}
\label{fig-coupled-representation}
\end{figure}

Now we can describe the coupled representations of $U(N)$.  The extended Young diagram for a coupled representation $R$ is built from 
two Young diagrams $R_+$, $R_-$, just by putting boxes in the shape $R_+$ at the upper left, antiboxes in the shape of an upside-down version of $R_-$ at the lower right, and zero-length rows in between so that the total height of the diagram is $N$.  An example is shown in Figure \ref{fig-coupled-representation}.  (Note that this procedure only makes sense for sufficiently large $N$, namely, $N$ has to be greater than the combined number of rows in $R_+$ and $R_-$.  We consider coupled representations for which one of $R_\pm$ has more than $\half N$ rows to be undefined.)
We write the coupled representation $\cR = R_+ \overline{R_-}$.

We will also need a slight generalization of this construction:  denote by $R_+ \overline{R_-} [l]$ the representation obtained by
tensoring $R_+ \overline{R_-}$ with $l$ powers of the determinant representation of $U(N)$.  This is equivalent to shifting
the lengths of all rows by $l$.

The representations $R_+ \overline{R_-}[l]$, where $R_+$ and $R_-$ are Young diagrams with $\le \half N$ rows, 
are a basis for the representation ring of $U(N)$ (at least for $N$ even.)  Another such basis would be obtained by
taking instead $R_+ \otimes \overline{R_-}$.  The two are not the same, although $R_+ \overline{R_-}$ is the principal component
of $R_+ \otimes \overline{R_-}$; the relation between the two is given by
\begin{equation} \label{tensortocoupled}
R_+ \otimes \overline{R_-} = \bigoplus_{S_\pm} \left[ \sum_{S'} N^{R_+}_{S_+ S'} N^{R_-}_{S_- S'} \right] S_+ \overline{S_-}
\end{equation}
(so long as $R_\pm$ each have $\le \half N$ rows; otherwise the right side would include $S_+ \overline{S_-}$
where one of $S_\pm$ has more than $\half N$ rows, which we have not defined.)
Here $S'$ and $S_\pm$ run over all (ordinary) Young diagrams.  Note that the only $S_+$ that 
contribute are ones which are contained in $R_+$, and 
similarly for $S_-$, so the sum in \eqref{tensortocoupled} is finite.  It gives the decomposition of $R_+ \otimes \overline{R_-}$ into irreducibles.

We will also need the inverse of \eqref{tensortocoupled}.  
To get it, we use the fact that the sum over $S'$ can be undone by summing
over another auxiliary Young diagram $R'$, using formula \eqref{ghost-cancellation}, here in the form
\begin{equation} \label{master-inverse-formula}
\sum_{R',S'} (-)^{\abs{R'}} N_{B_+ R' S'}^{A_+} N_{B_- R'^t S'}^{A_-} = \delta_{B_+}^{A_+} \delta_{B_-}^{A_-}.
\end{equation}
Applying this to \eqref{tensortocoupled} one obtains
\begin{equation} \label{coupledtotensor}
S_+ \overline{S_-} = \bigoplus_{R_\pm} \left( \sum_{R'} (-)^{\abs{R'}} N^{S_+}_{R_+ R'} N^{S_-}_{R_- R'^t} \right) R_+ \otimes \overline{R_-}.
\end{equation}
Again here, $R'$ and $R_\pm$ run over all ordinary Young diagrams, but only those $R_\pm$ which are contained
in $S_\pm$ contribute, so the sum is finite.

One can also rewrite \eqref{tensortocoupled} in terms of characters as
\begin{equation} \label{tensortocoupled2}
\sum_{R_\pm} \Tr_{R_+ \otimes \overline{R_-}}(U) \schur_{R_+}(V_+) \schur_{R_-}(V_-) = \sum_{S_\pm} \left( \sum_{S'} \schur_{S_+ \otimes S'}(V_+) \schur_{S_- \otimes S'}(V_-) \right) \Tr_{S_+ \overline{S_-}} (U),
\end{equation}
and \eqref{coupledtotensor} as
\begin{equation} \label{coupledtotensor2}
\sum_{S_\pm} \Tr_{S_+ \overline{S_-}}(U) \schur_{S_+}(V_+) \schur_{S_-}(V_-) = \sum_{R_\pm} \left( \sum_{R'} (-)^{\abs{R'}} \schur_{R_+ \otimes R'}(V_+)  \schur_{R_- \otimes R'^t}(V_-) \right) \Tr_{R_+ \otimes \overline{R_-}}(U).
\end{equation}

It is useful to know how to decompose the Casimir operators for $\cR = R_+ \overline{R_-} [l]$, 
\begin{align}
C_1(R_+ \overline{R_-} [l]) &= \abs{R_+} - \abs{R_-} + Nl,  \label{c1-split} \\ 
C_2(R_+ \overline{R_-} [l]) &= \kappa_{R_+} + \kappa_{R_-} + N (\abs{R_+} + \abs{R_-}) + 2l ( \abs{R_+} - \abs{R_-} ) + N l^2. \label{c2-split}
\end{align}
Here we introduced
\begin{equation}
\kappa_R = \sum r_i^2 - \sum c_i^2,
\end{equation}
where $r_i$ are the lengths of the rows of the Young diagram $R$ and $c_i$ are the lengths of the columns.
(So $\kappa_R = -\kappa_{R^t}$, where $R^t$ denotes the transpose of the diagram.)

\subsection*{Branching rules}

To understand the behavior of Yang-Mills theory when some eigenvalues are frozen, we need to understand
the branching rules for coupled representations:  how does a coupled representation of $U(N)$ decompose
under $U(N) \to U(N_1) \times U(N_2)$, for $N_1 + N_2 = N$?  In this section we will give formulas for these
branching rules.

We begin with the case of a representation $\cR$ of $U(N)$ which is given by an ordinary Young diagram,
$\cR = R$ (i.e. it can
be constructed using only fundamental indices, without the need for antifundamentals.)  In this case
the branching rule is well known (it is given e.g. in \cite{MR1354144} in the language of Schur functions),
\begin{equation} \label{branching-rule-fundamentals}
R \to \bigoplus_{R_1, R_2} N^R_{R_1 R_2} [R_1, R_2].
\end{equation}
Here $R_1$ and $R_2$ run over all Young diagrams (but of course the coefficient $N^R_{R_1 R_2}$ is only nonzero if $\abs{R_1} + \abs{R_2} = \abs{R}$.)  The same rule holds for representations $\overline{R}$ constructed only from antifundamentals.  Combining these
two rules we can find the branching rule for tensor products,
\begin{equation} \label{branching-rule-tensor-products}
R_+ \otimes \overline{R_-} \to \bigoplus_{R_{1\pm}, R_{2\pm}} N^{R_+}_{R_{1+} R_{2+}} N^{R_-}_{R_{1-} R_{2-}} [R_{1+} \otimes \overline{R_{1-}}, R_{2+} \otimes \overline{R_{2-}}].
\end{equation}
Now we can convert \eqref{branching-rule-tensor-products} into a branching rule for coupled representations.
We start with a coupled representation $R_+ \overline{R_-}$, apply \eqref{coupledtotensor} to write it in terms of tensor products, then apply \eqref{branching-rule-tensor-products} to decompose it under $U(N_1) \times U(N_2)$, then apply \eqref{tensortocoupled} to write the $U(N_2)$ part in terms of coupled representations again.  This leads straightforwardly to
\begin{equation}
R_+ \overline{R_-} \to \bigoplus_{A_\pm, Q_\pm} \left( \sum_{S',A'} (-)^{\abs{S'}} N^{R_+}_{A_+ Q_+ S' A'} N^{R_-}_{A_- Q_- S'^t A'} \right)  [Q_+ \otimes \overline{Q_-}, A_+ \overline{A_-}].
\end{equation}
But using \eqref{ghost-cancellation} the sums over $S'$ and $A'$ cancel one another, and
we are left with
\begin{equation} \label{branching-rule-coupled}
R_+ \overline{R_-} \to \bigoplus_{A_\pm, Q_\pm} \left( N^{R_+}_{A_+ Q_+} N^{R_-}_{A_- Q_-} \right)  [Q_+ \otimes \overline{Q_-}, A_+ \overline{A_-}].
\end{equation}
Note that for this formula to make sense we need that the $A_+ \overline{A_-}$ that appear are all well defined, which
requires that $R_\pm$ are shorter than $\half N_2$ rows.

Tensoring with powers of the determinant representation is also straightforward --- writing this representation $[1]$, 
it has the simple branching rule $[1] \to \left[[1],[1]\right]$.  This leads to
\begin{equation} \label{branching-rule-coupled-u1}
R_+ \overline{R_-}[l] \to \bigoplus_{A_\pm, Q_\pm} \left( N^{R_+}_{A_+ Q_+} N^{R_-}_{A_- Q_-} \right)  \left[ Q_+ \otimes \overline{Q_-} \otimes [l], A_+ \overline{A_-}[l] \right].
\end{equation}
The form of \eqref{branching-rule-coupled-u1} that we
ultimately use will be, when $\cR = R_+ \overline{R_-}[l]$ and $\cA = A_+ \overline{A_-}[l]$,
\begin{equation} \label{branching-rule-coupled-characters}
\Tr_{\cR / \cA} (e^{i\phi}) = \schur_{R_+ / A_+} (e^{i\phi}) \schur_{R_- / A_-} (e^{-i\phi}) \det(e^{i l \phi}).
\end{equation}
Here $e^{i \phi} \in U(N_1)$, and we emphasize again that \eqref{branching-rule-coupled-characters} holds only when $R_\pm$ are shorter than $\half N_2$ rows.

\subsection*{Factorization of $S_{\cP\cQ}$}

In order to understand the large-$N$ factorization of the $q$-deformed Yang-Mills with insertions,
we need to study the properties of the modular matrix $S_{\cP \cQ}$ of the
$U(N)$ Chern-Simons theory in the case where $\cP$ and $\cQ$ are coupled representations,
\begin{align}
\cP &= P_+ \overline{P_-}[l], \\ 
\cQ &= Q_+ \overline{Q_-}[m].
\end{align}
The most naive expectation would be that $S_{\cP\cQ}$ would be factorized into a piece
depending on $P_+, Q_+$ and a piece depending on $P_-, Q_-$.  
As we will show below, the correct formula is a sum of such terms,
\begin{multline} \label{spq-factorized}
S_{\cP\cQ} = M(q) \eta(q)^N 
q^{-\half (\kappa_{Q_+} + \kappa_{Q_-})  + (-\half N - m) \abs{P_+} + (-\half N + m) \abs{P_-} + (-\half N - l) \abs{Q_+} + (-\half N + l) \abs{Q_-} - 2lmN} \times \\
\sum_{R'} (-)^{\abs{R'}} q^{N\abs{R'}} C_{P_+ Q^t_+ R'} C_{P_- Q^t_- R'^t},
\end{multline}
where $C$ is the topological vertex of \cite{Aganagic:2003db} (in canonical framing.)
This formula was already obtained in \cite{Aganagic:2004js}, in the special case $P_\pm = 0$,
by direct computation using results from \cite{Iqbal:2004ne}.  

Here we will give a physical argument which explains
the reason for the factorization in the more general case of arbitrary $P_\pm$ and $Q_\pm$.\footnote{Our argument is not completely
rigorous, but we hasten to add that the final result has been checked on a computer for a variety of representations $P_\pm$
and $Q_\pm$.}  We restrict to the
case $l = m = 0$, i.e. $\cP = P_+ \overline{P_-}$ and $\cQ = Q_+ \overline{Q_-}$; the dependence on $l$ and $m$
is easily restored using \eqref{s-matrix-formula} and \eqref{c1-split}.
The idea is to realize the left side of \eqref{spq-factorized} as the partition function of the A model
topological string on the resolved conifold 
$T^*S^3$, with $N$ branes wrapped on $S^3$ and four infinite stacks of non-compact branes.  
Via the geometric transition of \cite{Gopakumar:1998ki} this is equal to the partition function
of the A model on the deformed conifold $\OO(-1) \oplus \OO(-1) \to \C\PP^1$, including the four
infinite stacks of non-compact branes \cite{Ooguri:1999bv}.  The latter partition function can be computed
using the topological vertex techniques of \cite{Aganagic:2003db}, which (as we will see) gives the right side
of \eqref{spq-factorized}.\footnote{Although the geometry $\OO(-1) \oplus \OO(-1) \to \C\PP^1$ is also
considered in the main text, the role the topological string is playing here is quite different from the
way it appears there.  We are using it here only as an auxiliary tool to prove the factorization of $S_{\cP \cQ}$.}

\begin{figure}[t]
\centering
\psset{unit=0.25}
\input{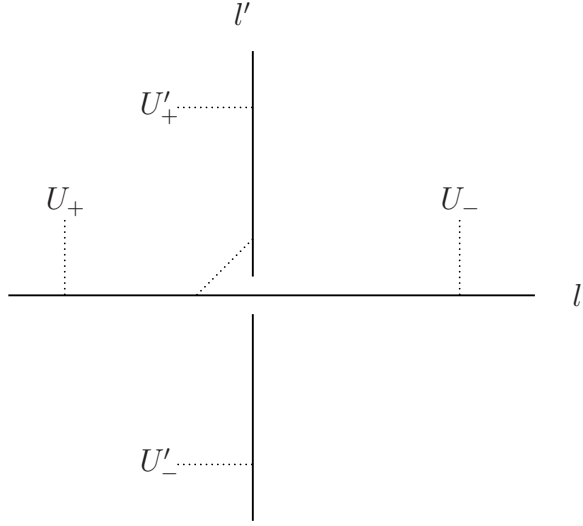}
\caption{A toric representation of the resolved conifold geometry $T^*S^3$, with the Lagrangian submanifold $S^3$ indicated, as well as four noncompact Lagrangian submanifolds with topology $S^1 \times \R^2$.  Each noncompact submanifold supports an infinite stack of branes with the $GL(\infty)$-valued complexified holonomy indicated.}
\label{fig-resolved}
\end{figure}

So we begin with the geometry $T^*S^3$.  As in \cite{Aganagic:2002qg}, we represent it as a $T^2 \times \R$
fibration over $\R^3$.  There are two lines $l$, $l'$ in $\R^3$ over which an $S^1$ of the $T^2 \times \R$ fiber degenerates, which are shown in Figure \ref{fig-resolved}.  Also shown in that figure is the Lagrangian submanifold $S^3$, which is a $T^2$
fibration over a line interval connecting $l$ and $l'$.  Finally, we also indicate four Lagrangian submanifolds of $T^*S^3$, constructed as described in \cite{Aganagic:2000gs}.  Each such submanifold has topology $S^1 \times \R^2$.

We will consider the topological A model on this geometry.  On each Lagrangian submanifold we place an infinite stack of A model D-branes.  There is then a $GL(\infty)$-valued complexified Wilson line on each stack, which couples to the open strings and thus enters the partition function.  We write these four Wilson lines $U_\pm$ and $U'_\pm$, as indicated in the figure.  We also include $N$ D-branes on the Lagrangian submanifold $S^3$.

The A model partition function in this geometry can be computed following \cite{Ooguri:1999bv,Aganagic:2002qg}.  The closed string partition function is essentially trivial --- it just gives an overall function of $q$, which is potentially ambiguous due to the non-compactness of the Calabi-Yau.  We set it to $1$ here.  The open string partition function receives contributions from annulus diagrams running along the lines $l$ and $l'$.  On each line there are three kinds of annuli which contribute:  one kind with the two boundaries on the two infinite stacks of branes, and two kinds with one boundary on an infinite stack and one boundary on the $N$ branes on $S^3$.

Integrating out the open string sector connecting the two infinite stacks, one gets a contribution to the partition function
\begin{equation} \label{pf-one}
\sum_R (-)^{\abs{R}} \schur_R(U_+) \schur_{R^t}(U_-),
\end{equation}
while the sectors connecting the infinite stacks to the $N$ branes on $S^3$ contribute operators
\begin{equation} \label{pf-two}
\left( \sum_{P_+} \schur_{P_+}(U_+) \Tr_{P_+}(V) \right) \left( \sum_{P_-} \schur_{P_-}(U_-) \Tr_{P_-}(V^\dagger) \right),
\end{equation}
with $V$ representing the holonomy around the $S^1$ where the annuli over $l$ meet $S^3$.

Combining \eqref{pf-one} and \eqref{pf-two}, one obtains for the open string contribution from the branes on $l$
\begin{equation} \label{open-string-contrib}
\sum_{R,P_\pm} (-)^{\abs{R}} \schur_{R \otimes P_+}(U_+) \schur_{R^t \otimes P_-}(U_-) \Tr_{\cP}(V).
\end{equation}
 Using the formula \eqref{coupledtotensor2}, \eqref{open-string-contrib} becomes
\begin{equation}
\sum_{P_\pm} \schur_{P_+}(U_+) \schur_{P_-}(U_-) \Tr_{\cP}(V).
\end{equation}
Similarly, from the branes on $l'$ we obtain
\begin{equation}
\sum_{Q_\pm} \schur_{Q_+}(U'_+) \schur_{Q_-}(U'_-) \Tr_{\cQ}(V'),
\end{equation}
where $V'$ is the holonomy on the $S^1$ where the annuli over $l'$ meet $S^3$.  Altogether, then, the contribution from open strings which involve the four infinite stacks of branes is
\begin{equation}
\sum_{P_\pm, Q_\pm} \left[ \schur_{P_+}(U_+) \schur_{P_-}(U_-) \schur_{Q_+}(U'_+) \schur_{Q_-}(U'_-) \right] \Tr_{\cP} (V) \Tr_{\cQ} (V').
\end{equation}
We view this $\schur_P(V) \schur_Q(V')$ as a product of Wilson line operators deforming the open string theory on the $N$ branes on $S^3$, namely the $U(N)$ Chern-Simons theory; these operators are arranged so as to 
give a Hopf link in $S^3$.  The Chern-Simons amplitude with this link inserted is simply $S_{\cP\cQ}$ \cite{Witten:1989hf}, so the topological string partition function is finally
\begin{equation} \label{top-string-cs}
\psi_\Top = \sum_{P_\pm, Q_\pm} \left[ \schur_{P_+}(U_+) \schur_{P_-}(U_-) \schur_{Q_+}(U'_+) \schur_{Q_-}(U'_-) \right] S_{\cP \cQ}.
\end{equation}

\begin{figure}[t]
\centering
\psset{unit=0.15}
\input{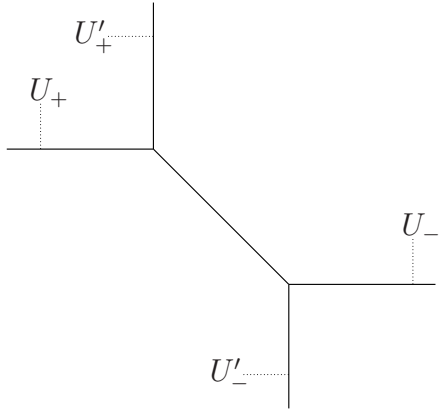}
\caption{The geometry of Figure \ref{fig-resolved} after the geometric transition.}
\label{fig-after-transition}
\end{figure}

To get the desired factorization of $S_{\cP\cQ}$ we now compute this partition function in another
way:  namely, we consider the geometric transition \cite{Gopakumar:1998ki} to the deformed conifold $\OO(-1) \oplus \OO(-1) \to \C\PP^1$, with the volume of $\C\PP^1$ given by $t = N g_s$, and with four infinite stacks of branes,
as shown in Figure \ref{fig-after-transition}.  Using the topological vertex of \cite{Aganagic:2003db}, the A model
partition function in this geometry can be computed as\footnote{One could determine the framing factors in \eqref{top-string-vertex} from first
principles; we determined them by requiring that the large $N$ limit of our factorization formula be correct.}
\begin{multline} \label{top-string-vertex}
\psi_\Top = M(q) \eta(q)^N \sum_{P_\pm, Q_\pm} \left[ \schur_{P_+}(U_+) \schur_{P_-}(U_-) \schur_{Q_+}(U'_+) \schur_{Q_-}(U'_-) \right] \times \\
q^{-\half \left(\kappa(Q_+) + \kappa(Q_-) + N(\abs{P_+} + \abs{Q_+} + \abs{P_-} + \abs{Q_-})\right)} \sum_{R'} (-)^{\abs{R'}} q^{N\abs{R'}} C_{P_+ Q^t_+ R'} C_{P_- Q^t_- R'^t}.
\end{multline}
More precisely, the factors $M(q) \eta(q)^N$ in \eqref{top-string-vertex} 
do not appear in \cite{Aganagic:2003db}, so they deserve some extra comment.  
The factor $M(q)$ is associated with the closed string sector; namely, in the large volume limit, it was shown in \cite{Gopakumar:1998ii,Faber:2000ma} that the closed A model partition function reduces to $M(q)^{\chi / 2}$ on a Calabi-Yau threefold with Euler characteristic $\chi$.  In the non-compact case
we are considering here $\chi$ is ambiguous, but the change in $\chi$ that occurs due to the geometric transition
would naturally be $\Delta \chi = 2$ (a 3-cycle gets replaced by a 2-cycle).  Thus, since we took $\chi = 0$ before the transition (we chose the closed string contribution in $\psi_\Top$ to be $1$), we should take $\chi = 2$ after the transition, which accounts for the factor $M(q)$.  The factor $\eta(q)^N$ is not as easily understood, but is presumably associated with the fact that $N$ D3-branes have disappeared in the transition; the same factor appeared in \cite{Saulina:2004da} associated to a single noncompact D3-brane.
Comparing \eqref{top-string-cs} and \eqref{top-string-vertex} one obtains the desired 
formula \eqref{spq-factorized}.

One can also compute a factorization formula for $1 / S_{0\cP}$, as was done in \cite{Aganagic:2004js}:
\begin{equation} \label{inverse-sp0-factorized}
\frac{1}{S_{0\cP}} = M(q)^{-1} \eta(q)^{-N} q^{\half N(\abs{P_+}+\abs{P_-})} \frac{\sum_{R} C_{P_+ 0R} q^{N\abs{R}} C_{P_- 0 R}}{C_{00P_+}^2{C_{00P_-}^2}}
\end{equation}
It would be interesting to know whether there is a physical argument for this factorization formula along the lines
of the argument given above for \eqref{spq-factorized}.

\subsection*{Factorization of $q$-deformed Yang-Mills amplitudes}

Now we are ready to consider the large $N$ factorization of the $q$-deformed Yang-Mills amplitudes with operator insertions.  We will approach the problem by factorizing each of the elementary ingredients from Appendix \ref{app-qym} separately.

In the large $N$ ('t Hooft) limit, a convenient basis for the Hilbert space $\HH$ is given by
the characters of the ``coupled representations'' which we introduced in Appendix
\ref{app-group-theory}; we write $\cR = \ket{R_+ \overline{R_-}[l]}$ for the states corresponding to the
coupled representations.  As was argued in \cite{Gross:1993hu}, these representations are
the only ones which contribute to the large $N$ amplitudes, to all orders in $1/N$; the reason is that they
are the only ones with $C_2(\cR) \sim N$.  All other representations are exponentially suppressed in the 't Hooft limit
by the factors $e^{- \half a g^2_\YM C_2(\cR) }$ which appear on a surface of area $a$ as in Appendix \ref{app-qym}
--- in the large $N$ limit they give contributions which are $\OO(e^{-N})$.

In this factorized basis, the ingredients of the amplitudes can be written as follows:

\medskip

\noindent {\bf The annulus.}  Using \eqref{annulus} together with \eqref{c1-split} and \eqref{c2-split}, we obtain easily
\begin{multline} \label{annulus-factorized}
\IP{R_{1+} \overline{R_{1-}}[l_1] \lvert A_a \lvert R_{2+} \overline{R_{2-}}[l_2]} = \delta_{R_{1+} R_{2+}} \delta_{R_{1-} R_{2-}} \delta_{l_1 l_2} e^{- a \half g^2_{\YM} N l^2} e^{i N a \theta_\YM l} \times \\
e^{- a \left(\half g^2_{\YM} (\kappa_{R_+} + N \abs{R_+} + 2l \abs{R_+}) - i \theta_\YM \abs{R_+} \right)}
e^{- a \left(\half g^2_{\YM} (\kappa_{R_-} + N \abs{R_-} - 2l \abs{R_-}) + i \theta_\YM \abs{R_-} \right)}. 
\end{multline}

\medskip
 
\noindent{\bf The Wilson line freezing operator.}  From \eqref{wilson-line-freeze-rep-basis} and \eqref{branching-rule-coupled-characters} 
we find
\begin{multline} \label{wilson-line-freeze-factorized}
\IP{R_{1+}\overline{{R}_{1-}}[l_1] \vert \delta_M(e^{i \oint A}, e^{i\phi}) \vert R_{2+} \overline{{R}_{2-}}[l_2]} = \\
\delta_{l_1,l_2} \left( \sum_{A_+} \schur_{R_{1+} / A_+}(e^{-i\phi}) \schur_{R_{2+} / A_+}(e^{i\phi}) \right) \left( \sum_{A_-} \schur_{R_{1-} / A_-}(e^{i\phi}) \schur_{R_{2-} / A_-}(e^{-i\phi}) \right).
\end{multline}

\medskip

\noindent{\bf The disc.}  From \eqref{disc} and \eqref{spq-factorized} this is
\begin{equation} \label{disc-factorized}
\Psi_0 = M(q) \eta(q)^N \sum_{l \in \Z, R_\pm} q^{-\half N (\abs{R_+}+\abs{R_-})} \left[ \sum_{R'} (-)^{\abs{R'}} q^{N \abs{R'}} C_{R_+ 0 R'} C_{R_- 0 R'^t}\right] \ket{R_+ \overline{{R}_-} [l]}.
\end{equation}

\medskip

\noindent{\bf The dual Wilson line freezing operator.}  
From \eqref{dual-wilson-line-freeze}, \eqref{spq-factorized} and \eqref{branching-rule-coupled-characters} we get
\begin{multline} \label{dual-wilson-line-freeze-factorized}
\Psi_\phi = M(q) \eta(q)^N \times \\
\sum_{l,m \in \Z, R_\pm, S_\pm} 
q^{-\half (\kappa_{S_+} + \kappa_{S_-})  + (-\half N - m) \abs{R_+} + (-\half N + m) \abs{R_-} + (-\half N - l) \abs{S_+} + (-\half N + l) \abs{S_-} - 2lmN} \times \\
\left[ \sum_{R'} (-)^{\abs{R'}} q^{N \abs{R'}} C_{R_+ S^t_+ R'} C_{R_- S^t_- R'^t}\right] \schur_{S_+}(e^{- i \phi}) \schur_{S_-} (e^{i \phi}) \det(e^{i m \phi}) \ket{R_+ \overline{{R}_-} [l]}.
\end{multline}

\medskip

\noindent {\bf The trinion (pair of pants).}  From \eqref{trinion} and \eqref{inverse-sp0-factorized} this is
\begin{equation} \label{trinion-factorized}
T = \sum_{l \in \Z, R_\pm} \left[ \sum_{R'} q^{\half N(\abs{R_+}+\abs{R_-})} \frac{C_{R_+ 0R'} q^{N\abs{R'}} C_{R_- 0 R'}}{C_{00R_+}^2{C_{00R_-}^2}} \right] \ket{R_+ \overline{R_-} [l]} \otimes \ket{R_+ \overline{R_-} [l]} \otimes \ket{R_+ \overline{R_-} [l]}.
\end{equation}

\renewcommand{\baselinestretch}{1}
\small\normalsize

\bibliography{physics}

\end{document}